\documentclass[runningheads]{llncs}

\newif\ifePrint
\ePrinttrue %\ePrintfalse%\ePrinttrue%

\usepackage[T1]{fontenc}
\usepackage[table,xcdraw]{xcolor}

\usepackage{graphicx}
\usepackage{xspace,soul}
\usepackage{appendix}

\usepackage{booktabs}
\usepackage{threeparttable}
\usepackage{cellspace, makecell, multirow}% for tables makegapedcells
\usepackage{amsmath}
\usepackage{amsfonts}
\usepackage{amssymb}
\usepackage{placeins}

\usepackage[nocompress]{cite}

\usepackage{xcolor,colortbl} % provides table cell color 
\usepackage{amsfonts}

\usepackage{tikz}
\usetikzlibrary{positioning}

\usepackage{circuitikz}
\usepackage[inline]{enumitem}
\usepackage{algorithm}
\usepackage{algorithmicx}
\usepackage[noend]{algpseudocode}

\usepackage[labelformat=simple]{subcaption}

\usepackage{mathtools}

\DeclarePairedDelimiter\largenorm{\lVert}{\rVert}%

\usepackage[colorlinks=true, allcolors=blue]{hyperref}
\usepackage{xspace}
\usepackage[misc,geometry]{ifsym}
\usepackage{fancyvrb}
\usepackage{fvextra}
\usepackage[framemethod=tikz]{mdframed}

\definecolor{cgrey}{gray}{0.65}
\definecolor{ccgrey}{gray}{0.85}
\definecolor{greenhtml}{HTML}{C6EFCE}
\definecolor{hyperplanegreen}{HTML}{b9d5cb}

\newcolumntype{g}{>{\columncolor{greenhtml}}c}
\newcolumntype{b}{>{\color{cgrey}}c}

\def\checkmark{\tikz\fill[scale=0.4](0,.35) -- (.25,0) -- (1,.7) -- (.25,.15) -- cycle;} 
\newcommand\cmark[1][]{%
    \tikz\fill[scale=0.35](0,.35) -- (.25,0) -- (1,.7) -- (.25,.15) -- cycle;}%
\newcommand\crossmark[1][]{%
  \tikz[scale=0.25,#1]{
    \fill(0,0)--(0.1,0) .. controls (0.5,0.4) .. (1,0.7)--(0.9,0.7) ..  controls (0.5,0.5) ..(0,0.1) --cycle;
    \fill(1,0.1)--(0.9,0.1) .. controls (0.5,0.3) .. (0,0.7)--(0.1,0.7) .. controls (0.5,0.4) ..(1,0.2) --cycle;
  }%
}

\newcommand{\dotp}[2]{{\langle #1, #2 \rangle}}
\newcommand{\norm}[1]{\| #1 \|}
\newcommand{\unitvec}[1]{\vec{ \hat #1}}%#{\mathbf{ \hat #1}}
\newcommand{\ndual}{N_{\textrm{dual}}}
\newcommand{\nexp}{N_{\textrm{exp}}}
\newcommand*\votes{\mathop{}\!\mathrm{votes}}
\newcommand*\CL{\mathop{}\!\mathrm{CL}}
\newcommand{\Y}{\checkmark}
\newcommand{\N}{\crossmark}
\usepackage{tikz}
\def\halfcheckmark{\tikz\draw[scale=0.3,fill=black](0,.35) -- (.25,0) -- (1,.7) -- (.25,.15) -- cycle (0.75,0.2) -- (0.77,0.2)  -- (0.6,0.7) -- cycle;}
\newcommand{\polynomialTime}{\small{\textit{poly}}}
\newcommand{\exponentialTime}{\small{\textit{exp}}}
\newcommand{\DeltaOn}{\mathrm{\Delta}_{\textrm{on}}}
\newcommand{\DeltaOff}{\mathrm{\Delta}_{\textrm{off}}}
\newcommand{\DeltaX}{\mathrm{\Delta}}
\newcommand{\eurocrypt}{Eurocrypt}
\newcommand{\asiacrypt}{Asiacrypt}
\newcommand{\crypto}{Crypto}
\newcommand{\xdual}{\vec x_{\textrm{dual}}} % alternatively $\mathbf x_d$, but that would be inconsistent with the images
\newcommand{\xdualhat}{\hat{\vec x}_{\textrm{dual}}}
\newcommand{\DistanceToToggle}{\textsc{DistanceToToggle}}

\newif\ifisdrafting
\isdraftingfalse
\mdfsetup{skipabove=0pt,skipbelow=0pt}

\usepackage{etoolbox}
\AtBeginEnvironment{tablenotes}{\footnotesize}

\begin{document}

\title{
% Polynomial Time Weight Extraction of ReLU-Based DNN Classifiers Given Only the Labels of Chosen Inputs \\
% \textcolor{blue}{\small{(voted for by: Adi, ... (please add your vote))}}\\
% \textcolor{blue}{OR} \\
Polynomial Time Cryptanalytic Extraction of Deep Neural Networks in the Hard-Label Setting \\
% \textcolor{blue}{\small{(voted for by: Anna, Isaac, Francisco ... (please add your vote))}} \\
% \textcolor{blue}{OR} \\
% Decision-Based Polynomial Time Cryptanalytic Extraction of Deep Neural Networks \\
% \textcolor{blue}{\small{(voted for by: Nicholas, ... (please add your vote))}}
}
% Crypto 20: "Cryptanalytic extraction of neural network models"
% EC24: "Polynomial Time Cryptanalytic Extraction of Neural Network Models"
% ASIACRYPT24: "Hard-Label Cryptanalytic Extraction of Neural Network Models" 

%Polynomial Time Cryptanalytic Extraction of Hard-Labeled Neural Network Models}%\thanks{Supported by organization x.}}

\titlerunning{Cryptanalytic Extraction of Deep Neural Networks}

\ifePrint
    \author{
    Nicholas Carlini\inst{1} \and 
    Jorge Chávez-Saab\inst{2} \and
    Anna Hambitzer\inst{2} \and
    Francisco Rodr\'iguez-Henr\'iquez\inst{2} \and 
    Adi Shamir \Letter\inst{3}
    }
\else
    \author{}
    % \author{
    % Nicholas Carlini\inst{1}\orcidID{\textcolor{red}{missing}} \and 
    % Jorge Chávez-Saab\inst{2}\orcidID{0000-0002-7006-1779} \and
    % Anna Hambitzer\inst{2}\orcidID{0000-0002-5357-832X} \and
    % Francisco Rodr\'iguez-Henr\'iquez\inst{2}\orcidID{0000-0002-5916-6625} \and 
    % Adi Shamir\Letter\inst{3}\orcidID{0000-0002-5422-905X}
    % }
\fi

\authorrunning{ }

\ifePrint 
    \institute{Google DeepMind 
    \email{nicholas@carlini.com}
    \and
    Cryptography Research Center, Technology Innovation Institute\\
    %\email{\{isaac.canales\}@tii.ae}
    \email{\{jorge.saab,anna.hambitzer,francisco.rodriguez\}@tii.ae}
    \and Weizmann Institute\\
    \email{adi.shamir@weizmann.ac.il}
    }
\else 
\institute{} 
\fi

\maketitle

\begin{abstract}
Deep neural networks (DNNs) are valuable assets, yet their public accessibility raises security concerns about parameter extraction by malicious actors. Recent work by Carlini et al. (\crypto'20) and Canales-Martínez et al. (\eurocrypt'24) has drawn parallels between this issue and block cipher key extraction via chosen plaintext attacks.
Leveraging differential cryptanalysis, they demonstrated that all the weights and biases of black-box ReLU-based DNNs could be inferred using a polynomial number of queries and computational time. However, their attacks relied on the availability of the exact numeric value of output logits, which allowed the calculation of their derivatives.
To overcome this limitation, Chen et al. (\asiacrypt'24) tackled the more realistic {\it hard-label scenario}, where only the final classification label (e.g., "dog" or "car") is accessible to the attacker.
They proposed an extraction method requiring a polynomial number of queries but an exponential execution time. In addition, their approach was applicable only to a restricted set of architectures, could deal only with binary classifiers, and was demonstrated only on tiny neural networks with up to four neurons split among up to two hidden layers. \\
This paper introduces new techniques that, for the first time, achieve cryptanalytic extraction of DNN parameters in the most challenging hard-label setting, using both a polynomial number of queries {\it and} polynomial time. We validate our approach by extracting nearly one million parameters from a DNN trained on the CIFAR-10 dataset, comprising 832 neurons in four hidden layers. Our results reveal the surprising fact that all the weights of a ReLU-based DNN can be efficiently determined by analyzing only the geometric shape of its decision boundaries.

\end{abstract}

\section{Introduction}
\label{sec:introduction}
% ================================================

%A Deep Neural Network (DNN) model $f_\theta$ is typically built by alternating between global linear operations (over $\mathbb{R}^{d_i}$ for various round dimensions $d_i$) and local nonlinear operations, such as the Rectifier Nonlinear Unit (ReLU) activation function, which is applied individually to each state component. Its parameters $\theta$ are generally obtained by collecting a huge set of training examples and modifying some initial $\theta_0$ by a lengthy sequence of gradient descent steps in order to minimize the loss of the DNN on the training examples. This process can require many months and cost millions of dollars, and thus a trained DNN can be a highly valuable asset. However, this asset is often made available for free through an oracle interface $\mathcal{O}$ that allows anyone to provide inputs to the DNN and obtain their corresponding outputs. The problem we address in this paper is whether it is possible to efficiently extract the exact values of all the parameters $\theta$ by interacting  with such a black box DNN.\commentFRH{Should we say first that we are addressing its hardest instance, the hard-label scenario?}
%
Deep Neural Networks (DNNs) have become ubiquitous in today’s technological landscape due to their ability to perform complex tasks such as image classification, speech recognition, natural language processing, and autonomous driving.

%ADI: I ELIMINATED THE PREMATURE USE OF f_THETA NOTATION HERE

The simplest form of a DNN consists of a series of fully connected hidden layers of neurons. Each neuron in the network performs a global linear operation (over $\mathbb{R}^{d_i}$ for various round dimensions $d_i$) followed by the parallel application of local nonlinear operations over $\mathbb{R}$, such as the Rectified Linear Unit (ReLU) activation function. The DNN's parameters are generally obtained by collecting a huge corpus of training examples and iteratively adjusting an initial set of parameters through a lengthy sequence of gradient descent steps aimed at minimizing the DNN's loss on the training examples. This process can take many months and incur costs in the millions of dollars, making a trained DNN a highly valuable asset. However, this asset is often made available for free through an oracle interface $\mathcal{O}$ that allows anyone to input data to the DNN classifier and receive the corresponding answer.   
% IT IS TOO EARLY TO DESCRIBE THIS LIMITATION BEFORE INTRODUCING THE CASE CONSIDERED BEFORE However, it is important to note that in a typical classification task, when given an input image, the oracle only provides the class label corresponding to the largest output logit, and does not usually provide the exact confidence score associated with this label.
  
%In this paper we investigate the research question of whether it is possible to efficiently extract the exact values of all the parameters $\theta$ by interacting  with a black-box DNN oracle that provides minimal information leakage in its outputs. (ADI: It is premature to talk about the research problem so early in the introduction)
%

%This model stealing problem has a rich history and had been dealt with in many papers over the last 30 years (as described in \autoref{sec:related_work}). Recently, cryptographic researchers had noticed the close similarity between the structure of DNN's and block ciphers (which are also composed of alternating layers of global linear operations and local nonlinear operations such as S-boxes). In both cases the nonlinear operations are publicly known, but the linear operations contain some secret elements (parameters $\theta$ in DNN's and round keys in block ciphers). The goal of the attacker is to find these secret elements by applying an adaptive chosen input attack. Not surprisingly, the best currently known model stealing attacks utilize ideas borrowed from the differential cryptanalysis of block ciphers.

\begin{figure}[!htb]
    \centering
    \includegraphics[width=1\linewidth]{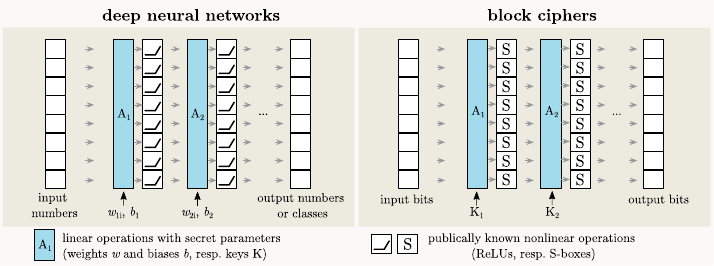}
    \caption{The similarity between DNN's and block ciphers.}
    \label{fig:fig-dnn-vs-block-cipher}
\end{figure}

The question of whether the DNN parameters can be determined by such an oracle access to its black box implementation has a rich history and had been addressed in numerous papers over the past~30 years (as detailed in \autoref{sec:related_work}). Recently, cryptographic researchers have noted the close similarities between the structures of DNNs and block ciphers, both of which consist of alternating layers of global linear operations and local nonlinear operations, such as ReLUs in the case of DNNs and S-boxes in the case of block ciphers (see \autoref{fig:fig-dnn-vs-block-cipher}).
In both primitives, the nonlinear operations are publicly known, while the linear operations involve secret elements — the parameters in DNNs, and the round keys in block ciphers. The goal of the attacker is to find these secret elements by applying an adaptive chosen input attack. Unsurprisingly, the most effective model stealing attacks currently leverage concepts from the differential cryptanalysis of block ciphers.

%A common taxonomy for attack scenarios on DNN classifiers was introduced in \cite{jagielski2020high}. It categorizes five distinct types of attacks, differentiated by the type of information provided by the black box DNN in response to each input query: (1) Just the most likely class label (this scenario is called the hard-label scenario), (2) the most likely class label and its probability score, (3) the top-k labels and probability scores, (4) all the labels and probability scores, and (5) the raw output of the DNN (i.e., all the logits before they are converted into a probability distribution). Clearly, scenario (1) is the hardest for the attacker, since for classifiers with two possible classes it provides only one bit of information per query, whereas scenario (5) is the easiest for the attacker since it provides the full numeric output of the DNN. 

A common taxonomy for attack scenarios on DNN classifiers was introduced in \cite{jagielski2020high}. They defined five distinct types of attacks based on the information provided by the black box DNN in response to each input query: 
\begin{enumerate*}[label=\textcolor{blue}{(S\arabic*)}]
  \item\label{item:hard-label-scenario} the most likely class label (referred to as the \emph{hard-label} scenario),
  \item  the most likely class label along with its probability score,
  \item the top-$k$ labels and their probability scores, 
  \item all labels and their probability scores,
  \item\label{item:easiest-scenario} the raw output of the DNN (i.e., all the logits before normalizing them into a probability distribution).
\end{enumerate*}
%(1) the most likely class label (referred to as the hard-label scenario), (2) the most likely class label along with its probability score, (3) the top-$k$ labels and their probability scores, (4) all labels and their probability scores, and (5) the raw output of the DNN (i.e., all the logits before normalizing them into a probability distribution). 
%Clearly, scenario (1) poses the greatest challenge for attackers, as it offers only one bit of information per query for classifiers with two possible classes, while scenario (5) is the least challenging, providing the complete numeric output of the DNN.
Scenario~\ref{item:hard-label-scenario} poses the greatest challenge for attackers, as it offers only one bit of information per query for classifiers with two possible classes, while scenario~\ref{item:easiest-scenario} is the least challenging, providing the complete numeric output of the DNN.

%Most previous attacks on black box DNN's had concentrated on the easiest attack scenario (5), and culminated with the attack of~\cite{canales2023polynomial} which requires a polynomial number of queries and a polynomial amount of time (as a function of the number of neurons). However, the problem of finding the best attack in the hardest scenario (1) was only dealt with in~\cite{chen2024} (after being first introduced in~\cite{Baum91}), which described an attack that uses a polynomial number of queries but an exponential amount of time.

Most previous attacks on black box DNNs have focused on the easiest scenario~\ref{item:easiest-scenario}, culminating in the work of \cite{canales2023polynomial}, which requires a polynomial number of queries and polynomial time (as a function of the number of neurons). However, the challenge of finding the most effective attack in the hardest scenario~\ref{item:hard-label-scenario} has only been tackled in \cite{chen2024}, over three decades after its initial introduction in 1991 in~\cite{Baum91}. Nonetheless, the solution in~\cite{chen2024} describes an attack that utilizes a polynomial number of \emph{queries} but requires exponential \emph{time}.\footnote{Moreover, Chen et al. achieved polynomiality only for stealing part of the parameters of the model, and only for a limited subset of DNN architectures.} This algorithmic limitation forced the authors of~\cite{chen2024} to report experiments on tiny networks with no more than four neurons split between at most two hidden layers.

\subsection{Why Previous Techniques Cannot Be Used}
\begin{figure}[hbt]
    \begin{subfigure}{0.0\textwidth}
    \phantomsubcaption\label{fig:intro-a}
    \phantomsubcaption\label{fig:intro-b}
    \phantomsubcaption\label{fig:intro-c}
    \end{subfigure}
    \centering
    \includegraphics[width=1\linewidth]{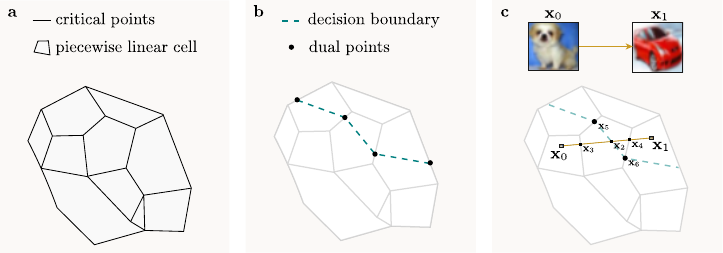}
    \caption{A schematic description of various attacks. 
    \textbf{a.}~Input space partition. % \autoref{fig:intro-a}. 
    \textbf{b.}~Decision boundary. % \autoref{fig:intro-b}. 
    \textbf{c.}~Path between two inputs. %\autoref{fig:intro-c}. 
    }
    \label{fig-introduction}
\end{figure}

To understand the difficulty posed by the harder scenario~\ref{item:hard-label-scenario}, consider the way most recent attacks operate, as depicted in \autoref{fig-introduction}. 
Any ReLU-based DNN represents a piecewise-linear mapping from real valued inputs to real valued outputs, which partitions the $d_0$ dimensional input space into a huge number of convex cells, as described in \autoref{fig:intro-a}. 
Within each cell, the mapping behaves as a linear transformation, and the boundaries between adjacent cells are $d_0-1$ dimensional hyperplanes. These hyperplanes represent the points where at least one of the ReLU inputs becomes zero, which we refer to as {\it critical points} following the terminology proposed in~\cite{carlini2020cryptanalytic} and adopted in~\cite{canales2023polynomial,chen2024}. Such ReLU flips change the behavior of the mapping from one linear mapping into a different one. Note that while the input/output mapping is always continuous, its derivatives become discontinuous at cell boundaries.
 
In addition to cell boundaries, we can also draw the decision boundaries between pairs of classes, whose points (which are called {\it transition points}) are the points at which the network's decision changes from one class to a different class.\footnote{Transition points are referred to as \emph{decision boundary points} in~\cite{chen2024}. For the sake of simplicity, we will ignore in this brief description the rare locations where there is a multi-way competition between three or more equal logits, where several decision boundaries intersect at the same point.}  Since the class decision is usually defined by the %\textcolor{blue}{\st{highest}} 
largest logit, decision boundaries are defined by linear inequalities between piecewise linear functions, and thus they are also piecewise linear $d_0-1$ dimensional hyperplanes which partition the $d_0$ dimensional input space into disjoint (not necessarily connected) regions, where each region represents the locations in which the hard-label assigned by the DNN is one of the classes. Note that decision boundaries can cut a single cell into two subcells which correspond to two different class decisions, even though the linear mapping in both subcells is the same. The only connection between cell boundaries and decision boundaries (which are defined by critical points and transition points, respectively) is that within each cell the decision boundary must be a flat hyperplane, which typically changes its orientation as it crosses into an adjacent cell, as depicted in \autoref{fig:intro-b}.

Most recent attacks have analyzed the output behavior as the input transitions along a straight line between two points in the input space, such as $x_0$ and $x_1$ in \autoref{fig:intro-c}. This path passes through various critical points such as $x_3$ and $x_4$. When dealing with the easiest attack scenario~\ref{item:easiest-scenario}, an attacker can precisely identify the location of such critical points, as the slope of the output changes abruptly. Each identified critical point offers valuable information about the value of some internal neuron during the evaluation of the DNN, by indicating that the input to its ReLU is zero at that point. This concept is akin to side-channel attacks that reveal internal values generated at various intermediate points during the encryption process, significantly reducing the complexity of key recovery. By collecting a sufficient number of such internal values, the attacker can recover the parameters of all the neurons by solving systems of linear equations in polynomial time.

The primary challenge in the hard-label attack scenario~\ref{item:hard-label-scenario} is the lack of access to the numeric values of the outputs, which prevents us from computing derivatives. Consequently, when we pass through critical points (such as points $x_3$ and $x_4$ in \autoref{fig:intro-c}),, we are unaware of this fact, as the DNN outputs "dog" for every input in the vicinity of $x_3$  and "car" for every input in the vicinity of $x_4$. The only event we can detect in this scenario is when we pass through the class transition point $x_2$ where the hard-label produced by the network changes from "dog" to "car". However, $x_2$ is not associated with any internal ReLU flip (in fact, it only reveals that the two top logits {\it at the end} of the DNN computation had become equal), and thus we no longer get the crucial side-channel information about values in the middle of the DNN computation which was revealed by crossing critical points in the easiest attack scenario~\ref{item:easiest-scenario}.

\subsection{Our Contributions}\label{sec:key-idea} 
The main contribution of this paper is to show that we can effectively replace the analysis of critical points in previous attacks by the analysis of class transition points, and thus recover all the DNN's secret parameters using a polynomial number of oracle queries and polynomial execution time by just analyzing the geometric shape of its decision boundary. 
In particular, we show that given any initial transition point, we can efficiently move along the decision boundary patch that contains it until this patch changes its orientation. This can happen only at a point which is simultaneously a transition point and a critical point, and thus we can indirectly sense the location of some critical points even though the attack scenario does not allow us to sense them directly. We call points which are both 
%\TODO{\st{decision points}} 
%\TODO
{transition points} and critical points {\it dual points}, and provide two examples of such dual points ($x_5$ and $x_6$) in \autoref{fig:intro-c}. Even though dual points are only an infinitesimally small subset of the set of all possible critical points, their analysis turns out to be a sufficient alternative to the analysis of critical points in previous attacks.

Our new attack follows the same strategy as outlined in \cite{carlini2020cryptanalytic} and further developed in \cite{canales2023polynomial} and \cite{chen2024}, but enhances them with novel techniques. 
The two main technical contributions of this paper are a new polynomial time algorithm for recovering the critical hyperplanes of all the neurons\footnote{We will refer to this process as \emph{signature recovery}, which will be discussed in \autoref{sec:signature_recovery}.}
and a new polynomial time sign recovery technique\footnote{That is, determining which side of the neuron's critical hyperplane provides positive values and which side provides negative values for the subsequent ReLU; this algorithm will be described in \autoref{sec:sign_recovery}} in the hard-label scenario. In particular, sign recovery was the main bottleneck in~\cite{chen2024}: the only solution the authors found for this problem was to perform an exponential time exhaustive search over all the possible sign combinations of all the neurons in the current layer.

It is important to note that our attack can fail in some extreme situations which are not likely to happen in normally trained networks (unless the network was adversarially generated to resist our attack). For example, if some neuron plays no role in forming the shape of the decision boundary, we will not be able to find its weights. Another rare possibility is that some system of $k$ random-looking linear equations in $k$ unknowns over floating point reals, generated by our attack, has a determinant that is exactly zero (which will make it impossible to recover any additional parameters at later layers).

Thanks to our enhanced signature and sign recovery algorithms in the hard-label scenario, we could demonstrate the practical applicability of our attack on a real DNN with~$935,370$ parameters that was trained to classify the CIFAR-10 classes. This DNN consists of an input layer of length 3072, three fully connected hidden layers with 256 neurons each, another fully connected hidden layer with 64 neurons, and finally 10 neurons that produce the 10 output logits. The best previous hard-label attack of \cite{chen2024} on this DNN would have required $2^{256}$ time, whereas our new method had succeeded in extracted all these parameters on a  standard multi-core server with GPU support. 

%%%%%%%%%%%%%%%%%%%%%%%%%%%%%%%%%%%%%%%%%%%%%%
%\vspace*{-\baselineskip}
\setlength{\tabcolsep}{6pt}
\begin{table}[htb]
\vspace*{-\baselineskip}
\begin{threeparttable}
\caption{\textbf{Comparison of Neural Network Extraction Methods.} Previous works successfully extracted either single-hidden-layer models or networks with very few hidden neurons, often requiring full access to floating-point outputs or resulting in exponential time costs. In contrast, our new attack demonstrates the ability to extract a multilayer neural network with approximately 1,000 hidden neurons under the most challenging hard-label setting. 
}
\label{tab:model-comparison}
\begin{tabular}{@{}p{\textwidth}@{}}
\centering
    %%%%% tabular part %%%%%%%%%
    \resizebox{\textwidth}{!}{%
    \footnotesize{
    \begin{tabular}{llc cc r}
    \toprule
    \multicolumn{3}{c}{\textbf{Neural Network}}
    & \multicolumn{2}{c}{\textbf{Parameter Extraction}} 
    & \\
    Architecture
    & Parameters
    & Hard-Label 
    & Signature
    & Sign
    & \textbf{Approach}  \\
    \cmidrule(rl){1-3} 
    \cmidrule{4-5}
    \cmidrule(lr){6-6}
    %%%
    10-20-20-1 
    & 620 
    & \N 
    & \cellcolor[HTML]{C6EFCE}{\color[HTML]{006100} \polynomialTime } % $2^{22}$ 
    & \cellcolor[HTML]{FFFFFF}{\exponentialTime}
    & ICML'20 \cite{RolnickK20} \\
    %%%
    10-20-20-1 
    & 620 
    & \N 
    & \cellcolor[HTML]{C6EFCE}{\color[HTML]{006100} \polynomialTime}%$2^{17.1}$ 
    & \cellcolor[HTML]{FFFFFF}{\exponentialTime} 
    & \crypto'20 \cite{carlini2020cryptanalytic} \\
    %%%
    784-128-1 
    & $\approx 0.1$M
    & \N 
    & \cellcolor[HTML]{C6EFCE}{\color[HTML]{006100} \polynomialTime}% $2^{21.5}$ 
    & \cellcolor[HTML]{FFFFFF}{\exponentialTime}
    & \crypto'20 \cite{carlini2020cryptanalytic} \\
    %%%
    3072-256$\times$8-10
    & $\approx 1.25$M
    & \N 
    & \cellcolor[HTML]{C6EFCE}{\color[HTML]{006100}  
  \textit{ \small (from \cite{carlini2020cryptanalytic})}} 
    & \cellcolor[HTML]{C6EFCE}{\color[HTML]{006100} \polynomialTime} 
    & EC'24 \cite{canales2023polynomial} \\ 
    %%%
    100-50-50-1
    & 7,650
    & \N 
    & \cellcolor[HTML]{C6EFCE}{\color[HTML]{006100} \polynomialTime}%  $2^{15.16}$  
    & \cellcolor[HTML]{C6EFCE}{\color[HTML]{006100} \polynomialTime}
    & \cite{Foerster24} \\ 
    \midrule
    %%%
   1024-2-2-1
    & 2,508
    & \cellcolor[HTML]{FFEB9C}{\color[HTML]{9C5700} \halfcheckmark} 
    & \cellcolor[HTML]{FFEB9C}{\color[HTML]{9C5700} \small{\textit{restr.}}}%$2^{22.38}$
    & \cellcolor[HTML]{FFFFFF}{\exponentialTime}
    & AC'24 \cite{chen2024} \\ 
    %%%
    3072-256$\times$3-64-10 
    & $\approx 0.9$M
    & \cellcolor[HTML]{C6EFCE}{\color[HTML]{006100} \Y} 
    & \cellcolor[HTML]{C6EFCE}{\color[HTML]{006100} \polynomialTime}%TODO  
    & \cellcolor[HTML]{C6EFCE}{\color[HTML]{006100} \polynomialTime}
    & {\it This work} \\ 
    \bottomrule 
    \end{tabular}%
    }}
    %%%%%%%%%%%%%%%%%%%%%%%%%%%%
    \end{tabular}
    \begin{tablenotes}
    \footnotesize \item[] \N:~In these works the attacker has full access to all the output logits in scenario~\ref{item:easiest-scenario}. \halfcheckmark:~Restricted hard-label scenario, e.g. only binary classifiers can be handled. \Y:~In these works the attacker has access only to hard-labels in scenario~\ref{item:hard-label-scenario}. 
    \polynomialTime: In these works, the signature, respectively sign recovery is done in polynomial time. \exponentialTime: In these works, sign recovery for most networks requires exponential time. \small{\textit{restr.}}: The signature recovery polynomiality is 
    only achieved for certain DNN architectures. 
    \end{tablenotes}
\end{threeparttable}
\vspace*{-\baselineskip}
\end{table}%

Table \ref{tab:model-comparison} compares our results to prior
extraction attacks both in the hardest hard-label setting~\ref{item:hard-label-scenario} and in the easiest logit-output setting~\ref{item:easiest-scenario}.

\subsection{Organization}
The remainder of this paper is organized as follows. In~\autoref{sec:related_work}, we provide a concise overview of the most relevant works in DNN parameter extraction within the black-box model along with the main precedents attacking the hard-label setting.
% \proposeRemove{In~\autoref{sec:preliminaries}, we outline the key definitions and assumptions that underpin this work, along with a formal problem statement.}
In~\autoref{sec:overview} we give an attack overview, and afterwards present a full description of the main components of our attack: dual point finding~(\autoref{sec:find-dual-points}), signature recovery~(\autoref{sec:signature_recovery}) and sign recovery~(\autoref{sec:sign_recovery}). All the practical experiments conducted in this study are described in~\autoref{sec:experiments}, while~\autoref{sec:conclusions} presents our concluding remarks. 

\FloatBarrier
% ================================================
\section{Related Work}
\label{sec:related_work}
% ================================================
\iffalse

Recent results on the problem:
- CJM20 - polynomial queries, but exponential in width
- Canales-Martinez 24 - fully polynomial
- Forster 24 (https://arxiv.org/pdf/2406.10011)
- Chen 24 - hard label! but back to exponential width

Our attack strictly dominates the scalability of prior work in the size
of models extracted.
%
Because of CJM20's exponential-time algorithm, they only recover
models with width up 20 neurons; in contrast, our attack recovers models
with over several hundred neurons wide and and over a thousand total hidden neurons.
%
TODO compare to C-M.
%
In contrast to Forster, 
%
Compared to the results of Chen \emph{et al.} who recover models
with \emph{four} hidden neruons, our attack can recover models with
a thousand hidden neurons and a several hundred thousand total parameters.\\

***\\
\fi

The problem of DNN model extraction was first studied in the 1990s~\cite{Baum90a,Baum91,BlumR93,HancockGM94}. 
In 1991, Baum presented an algorithm in~\cite{Baum90a,Baum91}  that could infer the Boolean function describing the model of a neural net algorithm in polynomial time. This was achieved by utilizing chosen inputs and querying the DNN as an oracle to obtain their labels. Baum demonstrated that his algorithm could provably achieve Probably Approximately Correct (PAC) learning in polynomial time for tiny networks consisting of up to four neurons, and he provided preliminary evidence suggesting that it could be extended to handle larger networks with up to 200 neurons. A few years after this research, Fefferman demonstrated in~\cite{Fefferman1994} that complete knowledge of all the (infinitely many) possible outputs from a sigmoid-based network uniquely determines its architecture and the weights of its neurons (up to some unavoidable symmetries). However, his technique did not provide an effective procedure for determining the actual network parameters. Another remarkable result on this topic was established by Blum and Rivest in 1993~\cite{BlumR93}. They examined a different scenario in which the attacker was given an adversarially chosen set of known inputs and their corresponding outputs (i.e., she could not choose her own queries). Their main result was that in this scenario,  determining whether there exists a corresponding two-layer, three-neuron DNN (with the sign activation function instead of a ReLU) is NP-complete.

%ADI: I WOULD ELIMINATE THE FOLLOWING PARAGRAPH SINCE ADVERSARIAL EXAMPLES ARE A COMPLETELY UNRELATED TOPIC. Academic and industrial research groups have shown increasing interest in this issue over the past decade. For instance, adversarial attacks, where imperceptible modifications to input images can lead to incorrect model predictions, have been studied in~\cite{PapernotMGJCS17,Carlini19,Park0M24}.

Since 2016, the extraction of DNN models from their black-box implementations has been extensively studied in~
\cite{tramer2016stealing,MilliSDH19,Reith0T19,jagielski2020high,abs-2105,Foerster24,MartinelliSGB24}. 
The main goal of this line of research is to accurately infer all secret weights of the neurons in the DNN, achieving sufficient numerical precision to ensure functional equivalence between the extracted model and the original DNN. Current state-of-the-art attacks in this domain can be found in~\cite{carlini2020cryptanalytic,canales2023polynomial,chen2024}.

In~\cite{carlini2020cryptanalytic}, the authors introduced several efficient techniques for recovering neuron weights and their associated biases with remarkable precision, a process they termed the \emph{neuron's signature recovery}. These methods operate with a polynomial number of queries and time complexity. However, this approach can only determine the neuron's signature up to a constant multiplier of unknown sign, thus necessitating an essentially exhaustive search for the correct signs, which has exponential time complexity. Consequently, the showcase examples of deep neural networks (DNNs) presented in~\cite{carlini2020cryptanalytic} involved relatively shallow networks. This limitation was addressed in~\cite{canales2023polynomial}, where the authors introduced novel techniques for recovering the missing neuron signs in polynomial time, making it possible to demonstrate attacks on significantly larger and deeper DNNs.

More recently, Foerster et al.~\cite{Foerster24} reported an end-to-end attack that effectively combines the signature recovery methods from~\cite{carlini2020cryptanalytic} with the sign recovery techniques from~\cite{canales2023polynomial}. They claim that their approach provides a more computationally efficient sign recovery process, resulting in significant speedups in the overall execution time of the attack. The authors note that, in this complete attack, most of the time is spent recovering signatures, with considerably less time devoted to extracting the signs.

Based on their experiments, they recommend brute-forcing the signs of particularly challenging neurons rather than spending excessive time trying to extract them. However, the DNNs targeted in~\cite{Foerster24} are notably shallow, consisting of no more than 100 neurons distributed across two hidden layers and a single logit output, which is not representative of real-world scenarios. For these relatively small networks, the authors demonstrate that it is feasible to exhaustively search the signs of a limited number of difficult-to-extract neurons. Nonetheless, it remains unclear how well this approach will scale to larger multi-output  DNNs with thousands of neurons spread across more than a handful of hidden layers.

All the attacks discussed in~\cite{carlini2020cryptanalytic,canales2023polynomial,Foerster24} fundamentally rely on the assumption that the DNN outputs confidence scores in the form of logits with 64-bit precision. This information allows attackers to compute derivatives and gradients with respect to the input, which is essential for conducting their signature and sign recovery procedures.

A more realistic scenario was studied in~\cite{chen2024}, where the authors explored the so-called hard-label setting. In this context, the DNN outputs only the label of the class with the highest confidence score, while the confidence score itself—an invaluable piece of information for the attacker—remains secret. Extracting DNN models in this hard-label setting presents significantly greater challenges than those based on the assumptions of previous studies. As a historical note, the concept of the hard-label setting was first proposed over three decades ago by Baum in~\cite{Baum90a,Baum91}, highlighting its lasting significance in this field and the persistent elusiveness of this problem that has remained unsolved until this paper.

The attack presented by Chen et al. in this hard-label scenario efficiently recovers signatures by exploiting the information provided by what the authors called \emph{decision boundary points}. However, their algorithm has several important technical limitations. In particular, their method achieves polynomial time complexity only for a very limited number of hidden layers and supports only single-bit binary DNN outputs. They did not succeed in developing an efficient algorithm for sign recovery, relying instead on a guessing approach, which incurs exponential time complexity. As a result, the signature recovery cannot be considered solved in the general case and the sign recovery problem remains the primary barrier to attacking larger and deeper networks in this setting. In fact, the DNNs successfully targeted in~\cite{chen2024} are restricted to those with a maximum of four neurons distributed across two hidden layers.

\FloatBarrier 
% ================================================
\section{Attack Overview}
\label{sec:overview}
% ================================================

\begin{figure}[!ht]
    \centering
    \includegraphics[width=\linewidth]{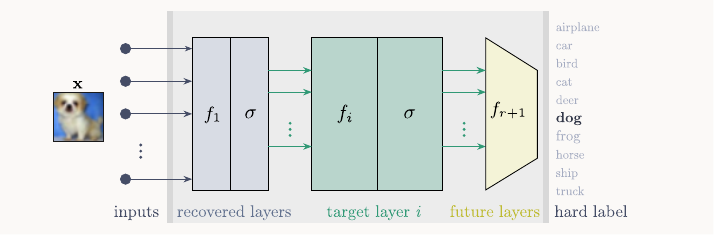}
    \caption{Representation of the DNN as a composition of the recovered layers $f_1,\ldots,f_{i-1}$, the current target layer $f_i$ and the non-recovered future layers $f_{i+1},\ldots, f_{r+1}$ (cf.~\autoref{def:DNN}).}
    \label{fig:Network}
\end{figure}

A deep neural network in the hard-label scenario takes inputs $\vec x$ and processes them layer-by-layer into a final output decision,~e.g.~"dog" (cf.~\autoref{fig:Network}).

\begin{definition}%[\cite{carlini2020cryptanalytic}]
\label{def:DNN}
% An \emph{$r$-deep neural network} $f$ is a function parameterized by $\theta$ that takes inputs from an input space $\mathbb{R}^{d_0}$ and returns values in an output space $\mathbb{R}^{d_{r+1}}$. The $i$-th \emph{fully connected layer} of a neural network is a function $f_i: \mathbb{R}^{d_{i-1}} \rightarrow \mathbb{R}^{d_i}$. The function $f$ is composed as a sequence of functions alternating between linear layers $f_i$ (of different dimensions $d_i$) and a nonlinear function (which acts component-wise) $\sigma$:
% \[ f = f_{r+1} \circ \sigma \circ \cdots \circ \sigma \circ f_{2} \circ \sigma \circ f_{1}. \]
An \emph{$r$-deep neural network} $f$ is a function parameterized by $\theta$ that takes inputs from an input space $\mathbb{R}^{d_0}$ and returns values in an output space $\mathbb{R}^{d_{r+1}}$. The function $f$ is composed as a sequence of functions alternating between linear functions $f_i: \mathbb{R}^{d_{i-1}} \rightarrow \mathbb{R}^{d_i}$, called \emph{fully connected layers}, and a nonlinear function $\sigma$ (acting component-wise):
\[ f = f_{r+1} \circ \sigma \circ \cdots \circ \sigma \circ f_{2} \circ \sigma \circ f_{1}. \]
\end{definition}

At a high level our attack extracts the parameters $\theta$ of a model
layer by layer, and the analysis of each layer consists of two steps:
\begin{itemize}
    \item \emph{Signature recovery} extracts the parameters of each neuron up to some unknown multiplicative factor. This determines the location of its critical hyperplane.
    \item \emph{Sign recovery} determines the side of the critical
    hyperplane in which the linear function of the neuron produces positive values which pass unchanged through the subsequent ReLU.
\end{itemize}

A full list of formal definitions can be found in~\autoref{sec:definitions}. Our terminology follows very closely the one first presented in~\cite{carlini2020cryptanalytic} and then adopted in more recent attacks~\cite{canales2023polynomial,chen2024}.

The main difference from previous attacks is that we no longer have direct access to critical points - we are only aware of class transition points. We will concentrate on the subset of transition points which are also critical points. These are the dual points, and we can find them by moving along decision boundaries until their local orientation changes~(detailed in \autoref{sec:find-dual-points}).

Since dual points are defined by the intersection of two locally linear $d_0-1$ dimensional hyperplanes (defined by the critical and transition conditions, respectively), they form a locally linear $d_0-2$ dimensional subspace $D$ which is associated with some neuron. Unfortunately, we can directly access only the transition hyperplane and not the critical hyperplane. We overcome this problem by computing the subspace $D$ by intersecting two adjacent decision boundary patches with different orientations (instead of intersecting a decision hyperplane and a critical hyperplane). In other words, once we discover some dual point $\xdual \in D$, we can explore the vicinity of  $\xdual$ in order to find the local orientations of the two decision boundary patches on its two sides, intersect them, and thus discover the whole subspace $D$ from a single point $\xdual$ in it. 
%By computing the intersection of these two $d_0-1$ dimensional patches, we can find their $d_0-2$ dimensional intersection, which is a locally linear subspace of dual points which are all contained in some $d_0-1$ dimensional critical hyperplane. Notice that while the original definition of dual points was as the intersection of transition and critical hyperplanes, in the hard-label setting we do not have direct access to critical points, so we find them instead by intersecting two adjacent decision boundary patches. 
Since $D$ is a $d_0-2$ dimensional subspace of the $d_0-1$ dimensional critical hyperplane of some neuron, knowledge of this $D$ makes it possible to find most (but not all) of the parameters of this neuron by just moving along the DNN's decision boundary and exploring its geometric shape. 

\begin{figure}[htb]
    \centering
    \includegraphics[width=\linewidth]{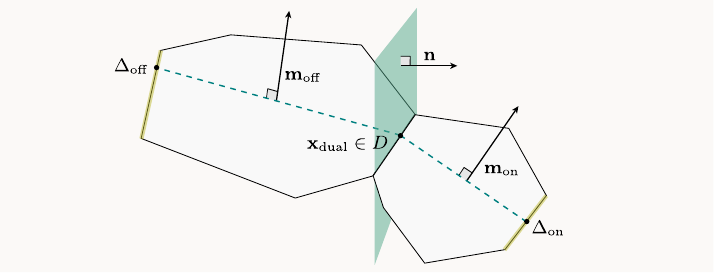}
    \caption{A green critical hyperplane of some neuron which changes the local orientation of two decision boundary patches on its two sides.}
    \label{fig-variables}
\end{figure}
A 3D depiction of this situation is depicted in \autoref{fig-variables}, where the vertical green 2D plane is the (unknown) critical subspace of some internal neuron, and the two 2D almost horizontal patches of the decision boundary on its two sides change their local orientation on the two sides of the critical plane. The intersection of the two 2D decision boundary patches yields a 1D line of dual points, which are all contained in the unknown critical plane. The only thing we do not observe about this critical plane is how it rotates around the known line of critical points within it.

To get a complete description of the neuron, we just have to find another instance of criticality of the same neuron in a different location of the input space. Our analysis is likely to yield a differently oriented subspace of dual points within it, and by combining the information they provide, we are likely to find the missing parameters of that neuron. Note that each partial information can be viewed as a set of linear equations, and thus determining whether two subspaces of dual points were produced by the same neuron, we simply have to check if the two systems of equations are consistent. If they are, their common solution is likely to provide a full description of this neuron, including its missing orientation around each one of the two subsets of dual points we have found within it.

The one remaining mystery about this neuron is which one of its two sides corresponds to positive inputs to its ReLU. This is referred to as the problem of finding {\it the sign of the neuron}, and without having this crucial information about all the neurons in the first $i-1$ layers of the DNN we will not be able to peel off the effect of these layers and proceed to the analysis of the $i$-th layer neurons, as depicted in \autoref{fig:Network}.

The attack presented in~\cite{canales2023polynomial} solved this problem in the easy scenario~\ref{item:easiest-scenario} by noticing that the output of the network tended to change faster (i.e., had a higher value for the absolute value of the slope) when the input $x$ moved from the side of negative ReLU inputs (where the output of this neuron was stuck at zero) to the side of positive ReLU inputs (where it was allowed to change and thus contributed to the overall change in the output). This effect could be maximized by wiggling the targeted neuron in a direction which is perpendicular to its critical hyperplane. While each individual test of slopes may give a wrong answer, we are likely to see a statistically significant difference between the speed of change of the outputs on the two sides of this neuron when we explore sufficiently many critical points belonging to the same neuron during a large number of traversals of the input space. Unfortunately, in the hard-label scenario~\ref{item:hard-label-scenario} considered in this paper we no longer have access to the numeric values of the output, and thus cannot compute the rate of change of these values. 

The way we solve the sign problem in the hard-label scenario is by exploiting a different type of measurement which is also strongly correlated to the speed of change of neuronal values, and can thus act as a proxy for the immeasurable output slopes. Consider \autoref{fig-variables} once again, where we are given some dual point $\xdual$. After finding the two $d_0-1$ dimensional decision boundary patches on its two sides, we can find the 
two dashed lines which start at $\xdual$, lie on the two decision boundary patches, and are perpendicular to $D$. Each one of these line segments ends when {\it another neuron} flips sides, and thus changes again the orientation of the decision boundary. We can now measure the two distances  $\DeltaOn$, $\DeltaOff$ until this happens on the two sides of the critical hyperplane. If this other neuron that flipped is at one of the later layers (compared to layer $i$ where the targeted neuron resides), then the speed at which its inputs change is affected by whether the targeted neuron is contributing to the change or not. This implies that typically (but not always), we expect decision boundary patches to be narrower on the positive side of the ReLU than on the negative side of the ReLU (i.e., that $\DeltaOn<\DeltaOff$). This is a measurable property even when the attacker has only access to the hard-labels assigned to inputs. In~\autoref{sec:experiments} we show experimentally that this statistical difference becomes very prominent once we test a sufficiently large number of dual points $\xdual$ which all lie on the critical hyperplane of the same targeted neuron. 

One important observation is that any particular patch of the decision boundary can be ended by flipping neurons which are located within the DNN either before or after the targeted neuron. Only neurons in later layers can possibly be affected by the question whether the targeted neuron's ReLU was on its positive or negative sides, and thus we should discard from our statistics any earlier layer neuron. Fortunately, we already know the signature of all these neurons, and we can test whether the dual subspace that {\it ends} the decision boundary patch yields a (partial) signature which is already known, and thus discard these cases.

\subsubsection{Adversarial Goals and Assumptions.}
According to our security model, the attacker can adaptively select queries $x$ as inputs to oracle $\mathcal{O}$, which returns the label of $x$ produced by the DNN $f_{\theta}$. The attacker's objective is to obtain the extracted parameters $\hat{\theta}$ which are the same as the original parameters ${\theta}$, up to roundoff errors produced by using finite precision real numbers\footnote{Note that by increasing the number of floating point bits we use by a factor of $k$, we can exponentially reduce these roundoff errors while increasing the time complexity of the arithmetic operations in our attack only by the polynomial factor of $k^2$.}, and up to unavoidable symmetries (such as permuting the order of internal neurons). In addition, we make the following typical assumptions regarding the capabilities of the attacker mirroring prior work on this topic \cite{carlini2020cryptanalytic}:
\textbf{Knowledge of the architecture}. The attacker has knowledge of the number of layers, the number of neurons in each layer, and the number of inputs and outputs in the network.
\textbf{Full-domain inputs}. The attacker can query arbitrary inputs from $\mathbb{R}^{d_0}$.
\textbf{Precise computations}. The DNN is specified and evaluated using a sufficiently high precision floating-point arithmetic. 
\textbf{Fully connected network and ReLU activation functions}. The network is comprised of fully connected layers and all activation functions are the ReLU function.
Compared to \cite{chen2024}, we do not assume that the classifier is binary, and can accommodate any number of classes (with multiple decision boundaries between them).

\FloatBarrier
% ================================================
\section{Dual Point Finding}
\label{sec:find-dual-points}
\begin{figure}
    \centering
    \includegraphics[width=1\linewidth]{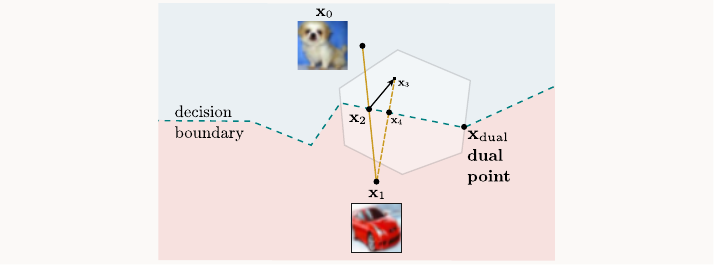}
    \caption{Visualization of our dual point finding algorithm.}
    \label{fig-find-dual}
\end{figure}
The value of critical points is well understood in the literature \cite{carlini2020cryptanalytic,canales2023polynomial},
because it is the location of these points that completely determine the
parameters of a neural network.
Unfortunately, without the ability to directly access the function $f(\cdot)$ that
returns the logits of the model,
it is not possible to identify general critical points.
Instead, we use dual points (cf.~\autoref{def:dual-point}): inputs that sit both at the
decision boundary and also on the critical hyperplane of some neuron. 
The finding of dual points is one key capability we used throughout our attack.
To identify dual points, we implement a simple algorithm, visualized in \autoref{fig-find-dual}: 

\paragraph{Step 1: find a point $x_2$ on the decision boundary.}
Initially, we sample two random inputs $x_0$ and $x_1$ that are labeled differently by the
neural network, i.e., $z(f(x_0)) \ne z(f(x_1))$.
We then perform binary search starting from these two points to find the exact location
of the decision boundary where $z(f(x_{2}))$ changes from one class to another.
Since we do this with binary search it is impossible to learn the \emph{exact}
location of the decision boundary, but with $\log_2 \epsilon$ queries we can recover a coordinate
satisfying $|f(x_{2})| < \epsilon$.
(We will also call this point $x_{\text{left}}$, because
later we will find a right point.)

\paragraph{Step 2: make a random excursion to $x_3$.}
We move away from $x_2$ in a random direction to reach a new point $x_3$. 

\paragraph{Step 3: find another point $x_4$ on the decision boundary.}
Evaluate the class $z(f(x_3))$. If the class corresponds to the one of $x_0$, find the transition point $x_4$ using a binary search between $x_3$ and $x_1$. If the class $z(f(x_3)) = z(f(x_1))$, find the transition point $x_4$ using a binary search between $x_3$ and $x_0$ instead. 

\paragraph{Step 4: move along the decision boundary until $\xdual$.}
Using the direction $dx=x_4 - x_2$ we can move to $x_\alpha = x_2 + \alpha \cdot dx$. For a small $\alpha$ the value $x_\alpha$ remains on the decision boundary.
But after some distance, eventually $x_\alpha$ will deviate away from the decision boundary.
Or, more accurately, the decision boundary will deviate away from $x_\alpha$.
Why is this?
Recall that ReLU neural networks divide the input space into piecewise linear regions.
Within any region, the decision boundary will behave completely linearly.
But once we cross from one linear region to another, 
the decision boundary will appear to \emph{bend},
because we are no longer in the same linear region.
The location of the bend is exactly the dual point which is both
on the decision boundary, and also on some neuron's critical hyperplane.

\paragraph{Step 5: re-locate the decision boundary.}
After crossing the critical hyperplane, we now project this point
back onto the decision boundary.
We refer to this point as $x_{\textrm{right}}$.
At this point, we have collected three points: $(x_{\textrm{left}}, \xdual$, and $x_{\textrm{right}})$.

\paragraph{Find the normal vector $m$ to the decision boundary.}
One final piece of information is necessary that will will use in several places: the normal vector to the decision boundary at the inputs $x_{\textrm{left}}$ and $x_{\textrm{right}}$.
Formally, the normal vector to the decision boundary $m$ is a
vector so that, for any input $x$ on the decision boundary and random vector $\epsilon$,
$x + (\epsilon - \text{proj}_m \epsilon)$ is too.
By viewing the normal vector this way, 
we can see how to compute its value through a series of queries.

To begin we choose (arbitrarily and without loss of generality) two unit-length
and orthogonal vectors $e_0$ and $e_1$.
We then choose a small constant $\alpha$,
and step in the direction $x' = x + \alpha e_0$.
Then, via binary search, we search for the smallest value of $\beta$ such that
$x^* = x' + \beta e_1$ is on the decision boundary.
Given the value of $\beta$ here, we can then compute $\beta \over \alpha$ as
the ratio between the first and second coordinates of the normal vector.
By repeating the above procedure for each of the remaining directions $e_2, e_3, \dots, e_{d_0}$,
we can completely reconstruct (up to scale) the normal vector $m$ to the decision boundary.

\FloatBarrier
% ================================================
%\section{Signature Recovery}
%\label{sec:overview}
% ================================================
%\commentFRH{@Nicholas please confirm if this structure fits you: The name of the section becomes 'Signature Recovery' and the overview of the attack is demoted to a subsection.}

% ================================================
\section{Signature Recovery}
\label{sec:signature_recovery}
% ================================================

% \commentFRH{
% Hi @Nicholas and @all. Thanks for this new version of this section and addressing my comments. Other than some small notations issues and the remaining TODOs, everything looks good now. 
% Some comments:
% \begin{enumerate}
%     \item We have not defined functionally equivalent
%     \item We have not defined signatures
% \end{enumerate}
% }

The first step in our attack recovers the parameters of each layer of the model up to
a real-value scalar per neuron.
Formally, the $i$-th fully connected layer of the neural network is $f_i: \mathbb{R}^{d_{i-1}} \rightarrow \mathbb{R}^{d_i}$, which is parameterized by the weight matrix $A^{(i)} \in \mathbb{R}^{d_i \times d_{i-1}}$ and the bias vector $b^{(i)} \in \mathbb{R}^{d_i}$~(cf.~\autoref{def:fi-Ai-bi}). 

\begin{definition}\label{def:signature}
The \emph{signature} of a neuron $\eta$ is equal to $\alpha \cdot A^{(i)}_\eta$,
where $\alpha$ is an an arbitrary rescaling of the corresponding parameter.
\end{definition}

It is easy to verify that positive constants can be pushed through the network arbitrarily 
(specifically: if all the weights and the bias of one neuron are multiplied by a constant $a>0$, 
and the corresponding inputs to every neuron on the next layer
are multiplied by $1 \over a$, then the model will behave identically). However, negative constants can not be pushed through
the model, since this causes neurons to flip from active to inactive
which changes the behavior of the model.
Therefore, our signature recovery attack mirrors the methodology of prior work \cite{carlini2020cryptanalytic}:
we search for a set of dual points,
and then use the locations of these dual points to determine the signatures on the first layer.

\subsection{Warm-up: Extracting the first layer}\label{sec:signature-first-layer}
The prior section developed an algorithm that allows us to efficiently identify a
vast set of dual points.
We now show how to use this information to recover the
normal vectors (up to sign and magnitude) to the neurons---and, as a consequence, the signatures
of all the neurons in the first layer of the neural network.

To begin, assume that we are given a dual (and thus critical) point $\xdual$ for some neuron $\eta$.
In prior work~\cite{carlini2020cryptanalytic}, 
computing the normal vector to the critical hyperplane induced by the neuron $\eta$ was possible directly via
finite differences, because they had the power to query the model $f$ at arbitrary
points $x \in \mathbb{R}^{d_0}$.
But now the only useful points our attack can sense are constrained to a piecewise linear $d_0-1$ dimensional subspace---the
set of points on the decision boundary of the neural network, and thus we must implement a slightly different attack,
formalized in Algorithm~\ref{alg:layer1_weights}.

\paragraph{Step 1: Compute the $(d_0-2)$-dimensional dual space in the vicinity of $\xdual$.}

\begin{definition}\label{def:dual-space}
    The \emph{dual space} $D^{\eta}$ of a neuron $\eta$ is the locally linear $d_0-2$
dimensional subspace which is the intersection between the
$d_0-1$ dimensional critical hyperplane for $\eta$ and the 
$d_0-1$ dimensional decision boundary in the vicinity of $\xdual$.
\end{definition}

This definition at first may not appear very useful:
we do not yet know the $d_0-1$ dimensional critical hyperplane,
and so how do we compute the intersection between this hyperplane
and the decision boundary?
But now let us make the following observation:
the dual space for a neuron $\eta$ is also
equal to the intersection of the two decision boundary hyperplanes
around the neuron, specifically, the ones at the points $x_{\textrm{left}}$ and $x_{\textrm{right}}$.
As a result, it is possible to compute the dual space (which has a
strong dependence on the parameters of the model) by making use of label-only oracle
queries to the model.

\paragraph{Step 2: Combine two dual spaces to recover the lost dimension.}\hspace{-.5em}\footnote{A simple 3D example of this process is to find a 2D plane from some 1D lines it contains. A single line does not fully define the plane, since it can rotate around it, but two different lines in it uniquely define the plane. Note that if the two lines are arbitrary lines in 3D space, they are extremely unlikely to belong to any common 2D plane, so we can cluster consistent lines that belong to the same neuronal plane.}
Suppose for the moment (and we will show how to do this next) that
we were given two different dual points, and their corresponding
dual spaces $D^\eta_0$ and $D^\eta_1$ for the same neuron.
Consider the quantity $A^{(1)}_j$: the $j$-th row of the first layer weight matrix;
put differently, this is the normal vector to the critical hyperplane.
Notice that each of these dual spaces satisfies $A^{(1)}_j \not\in D_0$
and $A^{(1)}_j \not\in D_1$
(because, by definition, the dual space contains the $d_0-1$ dimensional
critical hyperplane, which again by definition, does not contain the normal
vector to the hyperplane).

But also notice that, unless we are exceptionally unlucky, $D^\eta_0 \ne D^\eta_1$.
Therefore, by simply computing the union of these two spaces, we can recover
a $d_0-1$ dimensional space parameterized by its normal vector $n$.
And therefore, $A^{(1)}_j = n \cdot \alpha$ for some real valued (and possibly negative)
constant $\alpha$.

\begin{figure}[htb]
\vspace{-\abovedisplayskip}
\begin{minipage} [t]{0.42\linewidth}
\begin{algorithm}[H]%[htb]
    \footnotesize
    \caption{\footnotesize{\textsc{RecoverFirstLayerWeights}$(x_{\textrm{dual},0},\ x_{\textrm{dual},1})$}}
    \begin{algorithmic}[1]
        \Require $x_i$, dual points of the model
        \Ensure The parameters of the model $\alpha \cdot A^{(1)}_\eta$, or $\bot$ if the dual points are inconsistent.
        \State $D_i = \textsc{ComputeDualSpace}(x_i)$
        \State $D = D_0 \cup D_1$
        \If{$D = \mathbb{R}^{d_0}$}
            \State \Return $\bot$
        \Else
            \State Find $w$ such that \newline
                   \hspace*{1.15em} $\text{Span}(\{w\}) \cup D = \mathbb{R}^{d_0}$
            \State \Return $w$
        \EndIf
    \end{algorithmic}
    \label{alg:layer1_weights}
\end{algorithm}
\end{minipage}
\hfill
\begin{minipage}[t]{0.56\linewidth}
\begin{algorithm}[H]%[htb]
\footnotesize
\caption{\footnotesize{\textsc{CollectFirstLayerDualPoints}$(K)$}}
\begin{algorithmic}[1]
\Require $K$, number of dual points to search for
\Ensure Set of clusters of consistent dual points
\State $S \gets \emptyset$
\For{$i \gets 1$ to $K$}
    \State $\xdual \gets \textsc{FindRandomDualPoint}()$
    \State $S \gets S \cup \{\xdual\}$
\EndFor
\State $\text{clusters} \gets \emptyset$
\For{each $\xdual \in S$}
    \State $\text{matched} \gets \textbf{false}$
    \For{each $c \in \text{clusters}$}
        \State $r \gets \text{RandomElement}(c)$
        \If{\textsc{IsConsistent}$(r, \xdual)$}
            \State $c \gets c \cup \{\xdual\}$
            \State $\text{matched} \gets \textbf{true}$
            \State \textbf{break}
        \EndIf
    \EndFor
    \If{not $\text{matched}$}
        \State $\text{clusters} \gets \text{clusters} \cup \{\{\xdual\}\}$
    \EndIf
\EndFor
\State \Return $\{c \in \text{clusters} : |c| > 1\}$
\end{algorithmic}
    \label{alg:collectdual1}
\end{algorithm}
\end{minipage}
\end{figure}

\subsubsection{Filtering first-layer dual points.}

The final missing piece to our puzzle is to give an algorithm that
filters out the dual points that correspond to neurons
on the first layer from dual points that correspond to neurons on later layers.
We detail this algorithm in Algorithm~\ref{alg:collectdual1}, and describe it here.

Fortunately, this is straightforward and does not require any new work.
Suppose that we have two dual spaces $D_i$ and $D_j$ and want to know if they
correspond to the same neuron on the first layer.
Let us simply compute $E = D_i \cup D_j$ as the union of these two $d_0-2$
dimensional spaces.
We already know that $E$ is $d_0-1$ dimensional when both dual points
correspond to the same neuron on the first layer.
Let us now consider what happens if $D_i$ and $D_j$ correspond to different
neurons on the first layer, and after that, any neurons on later layers.

If $D_i$ and $D_j$ both correspond to different neurons on the first layer,
then we will have that $D_i^\perp = \text{span}(A^{(1)}_i, r_i)$ 
where $r_i$ acts as an arbitrary, essentially random vector (determined by later layers);
similarly $D_j^\perp = \text{span}(A^{(1)}_j, r_j)$.
Thus, because we assume that no two critical hyperplanes are exactly parallel,
we will have that $\text{span}(A^{(1)}_i, r_i) \cap \text{span}(A^{(1)}_j, r_j) = \emptyset$,
and therefore,
$D_i \cup D_j = \mathbb{R}^{d_0}$.
This allows us to distinguish dual points that share a neuron from those that do not.

\subsection{Extracting deeper layers}
Having shown how to recover the first layer, we now present our
general algorithm.
Without loss of generality, we assume that all previous layers of the network are extracted, and our goal is to extract the 
$i$-th layer.

\begin{definition}
The function that computes the first $i$ layers (up to and including $i$) of $f$ is called the input function and is denoted as $f_{1..i}$. In particular, $f = f_{1..r+1}$.
\end{definition}

Because we have extracted the parameters of the model up to layer $i$,
we can easily filter out and remove any dual points that belong to layers
$1$ through to $j$, by taking every dual point $\xdual$ and computing
$f_{1..j}(\xdual)$ for every $j <i$ and rejecting any dual point where
there exists some layer $j$ and neuron $\eta$ such that $f_{1..j}(\xdual)$ is at a critical point.

\subsubsection{Identifying layer-$i$ dual points}\label{sec:signature-layer-i-dual-points}

To start, we consider all dual points that have not already been found to be part
of some prior layer.
Now, we repeat an algorithm very similar to the algorithm from~\autoref{sec:signature-first-layer}.

\begin{definition}\label{def:nabla}
Let $f_{x_0;i}$ be the linear transformation satisfying $f_{x_0;i}(x_0) = f_i(x_0)$
and $\nabla f_{x_0;i}(x) \big|_{x=x_0} = \nabla f_i(x) \big|_{x=x_0}$.
\end{definition}

That is, $f_{x_0;i}$ is just the linear transformation that the function $f_i$ defines
within the linear region around the point $x_0$.

\begin{definition}
The hidden state at layer $j$ is the output of the function $f_{1..j}$, before applying the nonlinear transformation $\sigma$.
\end{definition}

For each dual point $\{\xdual{_{i}}\}$ and corresponding normal vector $\{n_i\}$,
we compute the hidden state $\xdualhat{_i} = f_{1..i}(\xdual{_{i}})$ and normal vector 
$\hat{n}_i = f_{\xdual{_i};i}(n_i)$.\footnote{Notice the reason we are using this function is because $n_i$ is not
an input to the neural network, but we just want to appropriately transform
the normal vector.}

By performing this linear transformation, we can now pretend that we are working
from within the space of the $i$-th hidden layer.
By doing this, it is now possible to check if two dual points are \emph{consistent}.

\begin{definition}
Two dual points $\xdual{_0}$ and $\xdual{_1}$ are \emph{consistent} if they occur because of the
same neuron $\eta$ on the same layer $i$ of the model.
\end{definition}

To check consistency, we leverage the fact that the union of the dual spaces for
dual points that share the same neuron will not have full rank,
whereas the dual spaces for dual points from different neurons will have full rank.

This allows us to repeat our prior algorithm identically;
with one key challenge.
Because we are working with ReLU networks, all inputs that are negative will be
set to zero.
Therefore, the linear transformation defined by
$f_{x_0;i}(x_0)$ will not be \emph{onto}---in fact, it will have a null space
roughly equal to $d_i/2$~\cite{canales2023polynomial}.

\begin{definition}
    A \emph{partial dual space} $D_{\text{partial}}$ is a subspace of the
    dual space $D$.
\end{definition}

What this means practically speaking is that when we compute the union of two
dual spaces, even if they correspond to completely different neurons, they will (with high probability) share some
coordinates which are identically zero.
Therefore, the correct measure to check if two dual points are consistent is not
if they have completely full rank, 
%but \emph{as full rank as they could have},given the number of shared coordinates where both dual points have dead neurons.
but rather whether they achieve the highest rank possible, given the number of shared coordinates where both dual points have dead neurons.

\begin{algorithm}[htb]
    \footnotesize \caption{\footnotesize{\textsc{IsConsistent}$(\xdual{_0}, \xdual{_1}, D_0, D_1)$}}
    \begin{algorithmic}[1]
        \Require $\xdual{_i}$ a dual point of the model, $S_i$ the corresponding dual space
        \Ensure True if the two dual points are on the same neuron's critical hyperplane, otherwise False.
        \State Let $\tilde{D}_j = f(\xdual{_i};i)(D_j)$
        \State Let $D = \tilde{D}_0 \cup \tilde{D}_1$.
        \State Let $X$ = \text{the number of neurons active in either $\xdual{_0}$ or $\xdual{_1}$.}
        \If{$\text{Rank}(D) = X$}
          \State \Return \text{No}
        \Else
          \State \Return \text{Yes} \Comment{Specifically, $Rank(D) < X$}
        \EndIf
    \end{algorithmic}
    \label{alg:layer_isconsistent}
\end{algorithm}

\subsubsection{Recovering layer-$i$ signatures}\label{sec:deeper-layer-signature}

Given a method to identify whether or not two layer-$i$ dual points correspond
to the same neuron, we can now finally recover the signatures for this layer
of the model.

\paragraph{Step 1: cluster dual points.}
Given all of the dual points we have found thus far,
we first cluster them together using the $\textsc{IsConsistent}$ function
from Algorithm~\ref{alg:layer_isconsistent}.
It is straightforward to implement an $O((\text{\# duals})^2)$
work algorithm to achieve this.
We begin by constructing a graph; each dual point corresponds to a node,
and we connect two dual points with an edge if they are consistent.

In an ideal world, this graph would be disconnected and have exactly $d_i$ cliques
of size greater than one, and all remaining nodes isolated from each other
(corresponding to neurons on deeper layers).
In practice, however, with low probability the $\textsc{IsConsistent}$ algorithm
will spuriously claim two neurons are consistent when in fact they are
not with low (e.g., $10^{-6}$) probability due to numerical instabilities.
To resolve this issue,
we instead search for \emph{maximal} clique,
because we find that spurious errors are typically independent;
this way, single edges that connect to cliques are automatically discarded.

\paragraph{Step 2: unify dual spaces to recover signatures.}
At this point in our algorithm we have one set of dual points per
neuron on the $i$-th layer of the model.
All that remains now is to recover the parameters for each of these neurons.

Fortunately, in~\autoref{sec:signature-first-layer} where we introduce our layer-$1$ attack,
we have already done almost all of the work necessary.
Only now, instead of collecting a pair of $d_0-2$ dimensional dual spaces
and unifying them together to obtain a $d_0-1$ dimensional critical hyperplane,
we now unify a larger set $\Omega(\log(d_i))$ dual points together.

The reason why we need a larger number of dual points per neuron is that
a small constant fraction of the neurons in any given layer will be dead \cite{canales2023polynomial}.
And so, in order to observe at least one non-dead input in each of the $d_i$
positions, we should expect to need to see $\Omega(\log(d_i))$ different examples
assuming a perfectly uniform distribution of active and inactive neurons.
If we collect our set of dual points and find that they do not have sufficient
diversity, then we simply return back to the first step of our attack and
collect more dual points.
\begin{algorithm}[htb]
    \footnotesize
    \caption{\footnotesize{\textsc{RecoverDeeperWeights}$(\{\xdual{_j}\}_{j=1}^n, \{D_j\}_{j=1}^n)$}}
    \begin{algorithmic}[1]
        \Require $\xdual{_j}$ a dual point of the model, $D_j$ the corresponding dual space
        \Ensure The parameters of the model $A^{(i)}_\eta$.
        \State \textbf{assert} all dual points $\xdual{_j}$ are (pairwise) consistent.
        \State Let $\tilde{D}_j = f(d_j;i)(D_j)$
        \State Let $D = \tilde{D}_0 \cup \tilde{D}_1 \cup \tilde{D}_2 \cup \dots \cup \tilde{D}_n$
        \If{$\text{Rank}(D) < d_0-1$}
            \State \Return $\bot$ \Comment{Insufficient data to reconstruct parameters.}
        \Else
            \State Find $w$ such that $\text{Span}(\{w\}) \cup D = \mathbb{R}^{d_0}$
            \State \Return $w$
        \EndIf
    \end{algorithmic}
    \label{alg:layerk_weights}
\end{algorithm}
A complete description of the algorithm for this attack is given in Algorithm~\ref{alg:layerk_weights}.

\FloatBarrier

% ================================================
\section{Sign Recovery}
\label{sec:sign_recovery}
% ================================================
%Recall that a neuron's signature is related to its actual weights by some arbitrary scaling factor $c$. As shown in \cite{canales2023polynomial}, the absolute value of $c$ is immaterial, since we can multiply all the weights of some neuron in layer $i$ by some positive multiplier and compensate for it by dividing all the coefficients that refer to this neuron at layer $i+1$ by the same positive scaling factor, which leaves the outputs of the DNN unchanged. However, finding the correct \emph{sign} of $c$ is crucial in order to recover a functionally equivalent model. Similarly to~\cite{canales2023polynomial}, we use a heuristic argument and develop a statistical test that is likely to identify the correct sign. \\
%
Recall that a neuron's signature is related to its actual weights by some arbitrary scaling factor $c$. Finding the correct \emph{sign} of $c$ is crucial in order to correctly recover the model. Similarly to~\cite{canales2023polynomial}, we use a heuristic argument and develop a statistical test that is likely to identify the correct sign.
We begin by providing the intuition behind our sign recovery technique and experimentally validating its key assumption (\autoref{sec:sign_recovery-intution}). Next, we present the technique in algorithmic form (\autoref{sec:sign_recovery-algorithm}).%, and conclude by introducing the confidence level $\CL$ for sign recovery (\autoref{sec:confidence-level}).  

% ================================================
\subsection{The Basic Technique}
\label{sec:sign_recovery-intution}
% ================================================
To correctly determine the sign of a target neuron, we require a property that differs measurably between the neuron's on- and off-sides.

%\subsubsection{Neuron Wiggle.} 
%\textcolor{magenta}{\st{At Eurocrypt'24,}} 
In \cite{canales2023polynomial} the authors demonstrated that a small perturbation (\textit{neuron wiggle}) in a carefully chosen direction is expected to change the target neuron's output by $\epsilon$, while affecting all other neurons in the same layer by about $\pm \frac{\epsilon}{\sqrt{d}}$, where $d$ is the target layer's width. Consequently, the output norms of the target layer, $||v_{\textrm{on}}||$ and $||v_{\textrm{off}}||$, differ between the on- and off-sides of the target neuron. Assuming that the neurons in the remaining layers behave randomly, the changes of the floating point values of the output logits also differ in their magnitude between the on- and off-side. This approach allowed the authors of~\cite{canales2023polynomial} to directly measure these infinitesimal changes in the output logits of the neural network.

\begin{figure}[!htb]
    \centering
    \includegraphics[width=1\linewidth]{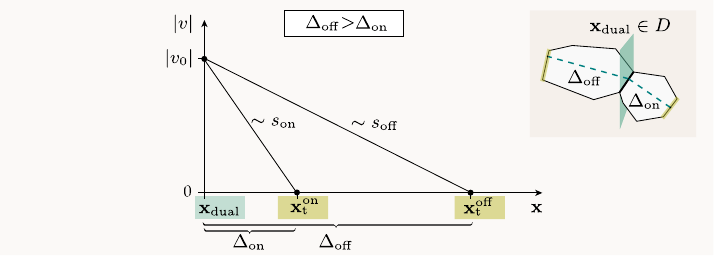}
    \caption{Assume a target neuron at a dual point $\xdual$, and a future neuron at initial value $v_0$. A higher average speed on the on-side $s_{\textrm{on}}$ compared to the off-side $s_{\textrm{off}}$ will result in a measurable difference $\DeltaOff > \DeltaOn$.}
    \label{fig-intuition}
\end{figure}

% %%%%%%%% ||vON|| vs ||vOFF|| %%%%%%%%%
% At Eurocrypt'24 it was shown by \cite{canales2023polynomial} that a {\it neuron wiggle} will change the current target neuron output by $\epsilon$, while changing all other neurons in the target layer only by $\pm \frac{\epsilon}{\sqrt{d}}$, where $d$ is the target layer width. 
% The corresponding target layer output norms $||v_{\textrm{on}}||, ||v_{\textrm{off}}||$ will be different on the on-, respectively off-side. 
% \cite{canales2023polynomial} could directly measure the effect of the different target layer output norms at the output-logits of the neural network. 
%
%%%%%% sON vs sOFF %%%%%%%%%
%\subsubsection{Sign Recovery without Logits.}
One key intuition for sign recovery without direct access to output logits is, that the average speed with which \emph{all} future neurons\footnote{A \emph{future neuron} means a neuron that is located in any of the layers $f_{i+1}, \ldots, f_{r+1}$ that \emph{follow} the current target layer $i$.} change their values depends on the target layer output norms. Accordingly, the speed with which the average future neurons change their value on the target neuron's on-side, $s_{\textrm{on}}$, is larger than that on its off-side, $s_{\textrm{off}}$, see also \autoref{fig-intuition}. 
The core assumption is that this speed difference correlates with the target layer's output norms, leading to the approximate magnitude estimates given in \autoref{eq:son} and \autoref{eq:soff}.
\begin{align}
s_{\textrm{on}} & \propto ||v_{\textrm{on}}|| \approx  \largenorm{\left(\begin{matrix}\pm\frac{\epsilon}{\sqrt{d}}, & \pm\frac{\epsilon}{\sqrt{d}}, & \ldots, & \epsilon &,\ldots, & \pm\frac{\epsilon}{\sqrt{d}}, & \pm\frac{\epsilon}{\sqrt{d}}\end{matrix} \right) }\label{eq:son} \\
s_{\textrm{off}} & \propto ||v_{\textrm{off}}|| \approx \largenorm{\left(\begin{matrix}\pm\frac{\epsilon}{\sqrt{d}}, & \pm\frac{\epsilon}{\sqrt{d}}, & \ldots, & 0 &,\ldots, & \pm\frac{\epsilon}{\sqrt{d}}, & \pm\frac{\epsilon}{\sqrt{d}}\end{matrix} \right)} \label{eq:soff} 
\end{align}
%
%%%%%% Delta ON vs Delta OFF %%%%%%%%%
%\vspace{0.3em} \noindent \textbf{Measuring $\DeltaOn$ vs. $\DeltaOff$.} 
Any neuron toggling point introduces bends in the decision boundary, which allows us to measure $\DeltaOn$ and $\DeltaOff$. If a future neuron is approaching its toggling point $\vec x_{\textrm{t}}$, a higher speed implies it will reach this point \textit{earlier} than if moving at a slower speed (see \autoref{fig-intuition}). The ratio between the two very rough theoretical estimates of speed given above is about $\sqrt{2} \approx 1.4$, while actual experiments yield an average speed ratio of about $1.2$. This significant ratio of distances $\DeltaOff / \DeltaOn$ (which does not depend on the dimension $d$ and is expected to remain roughly the same at different layers) is easily measurable. By running a small number of experiments and taking their majority vote (i.e., assigning a minus sign to the side of its critical hyperplane which tends to produce longer distances) we are extremely likely to find the correct sign of the neuron.

Note that in the case of perfect control only the target neuron will change by a value $\epsilon$ on the on-side while all other neurons in the target layer will change by $0$, resulting in $||v_{\textrm{on}}||/||v_{\textrm{off}}||=\infty$. In this extreme case of perfect control it is obvious that $\DeltaOff > \DeltaOn$ since there will never be a future toggle on the off-side.\footnote{If we move too far away from $\xdual$ there will however be a past layer toggle at some point. We discuss later in detail how these are handled.} 

\subsubsection{Experimental Verification of the Basic Technique} 
To experimentally verify our technique in terms of the level of control ($||v_{\textrm{on}}||/||v_{\textrm{off}}||$) and the average future neuron speed ($s_{\textrm{on}}$ and $s_{\textrm{off}}$), we record these values in a white-box setting for the network described in \nameref{sec:our-network}. Details of this network will be provided in the experimental section; here, we note that it contains four hidden layers, with 256 neurons in hidden layers~1 to~3, and~64 neurons in hidden layer~4.
For a large number of target neurons in every hidden layer, we examine the speed of all future neurons and the level of control during our sign recovery attack\footnote{The construction details of the neuron wiggle for the attack are discussed in \autoref{sec:sign_recovery-algorithm}.} across several dual points. At each dual point, the speed of each future neuron is calculated as $s_{\textrm{on, off}} = |v\left(\xdual\right) - v\left(\vec{x}_{\textrm{t}}\right)| / \DeltaX_{\textrm{on, off}}$. This represents the change in the neuron's value before its ReLU, divided by the distance $\DeltaX_{\textrm{on, off}}$ to the first future neuron toggling point on the on- or off-side.

%%%%%%%%%%%%%%%%%%%%%%%%%%%%%%%%%%%%%%%%%%%%%%
% The data in this table has been generated using 
% /home/anna/file-system/pure/share/users/anna_pub/2024-09_DETI-II/D/02_CheckSpeedIntuition.ipynb
\setlength{\tabcolsep}{6pt}
%\vspace*{-\baselineskip}
\begin{table}[htb]
\vspace*{-2\baselineskip}
\begin{threeparttable}
\caption{Whitebox-verification on the intuition for the sign-attack across target neurons in \nameref{sec:our-network}. %\TODO{There is one neuron not yet analyzed in layer 1, and 5 neurons (3,13,46,53,61) not yet analyzed in layer 4.}
}
\label{tab:intuition}
\begin{tabular}{@{}p{\textwidth}@{}}
\centering
    %%%%% tabular part %%%%%%%%%
    %\resizebox{\textwidth}{!}{%
    \footnotesize{
    \begin{tabular}{ll cc ccc}
    \toprule
     & 
     & \multicolumn{2}{c}{\textbf{Average target neuron}} 
     & \multicolumn{3}{c}{Worst target neuron} \\
         Layer  
        & Neurons 
        & $||v_{\textrm{on}}||/||v_{\textrm{off}}||$  
        & $s_{\textrm{on}}>s_{\textrm{off}}$ 
        & $n_{\textrm{ID}}$ 
        & $||v_{\textrm{on}}||/||v_{\textrm{off}}||$ 
        & $s_{\textrm{on}}>s_{\textrm{off}}$ \\
        \cmidrule(rl){1-2} \cmidrule{3-4} \cmidrule(lr){5-7} 
        %\cmidrule(lr){7-8}
    1 & 256       
    & $\infty$ & 100\% 
    & - & - & -  \\ 
    2 & 256       
    & $(1.20\pm 0.04)$ & $(96\pm5)\%$ 
    & 170 & 1.07 & 64\% \\
    3 & 256       
    & $(1.21\pm 0.04)$ & $(82\pm7)\%$ 
    & 226 & 1.19 & 59\% \\
    4 & 64 
    & $(3.2 \pm 0.8)$ & 100\% 
    & - & - & -  \\ 
    \bottomrule 
    \end{tabular}%
    }%}
    %%%%%%%%%%%%%%%%%%%%%%%%%%%%
    \end{tabular}
    \begin{tablenotes}
    \footnotesize \item[]
    %\footnoterule
    %\item[\textdagger] ...
    \end{tablenotes}
\end{threeparttable}
\vspace*{-2\baselineskip}
\end{table}

\autoref{tab:intuition} shows that the intuition on the future neuron speed being larger on the on- than the off-side of the neurons is true for all the   $(3\times256+64=832)$ target neurons in the network. the average ratio is about 1.2, and even for the worst neurons the ratio is above 1.07 (which will simply require a larger number of corresponding dual points to get a reliable result).

%For each target neuron in each hidden layer, we analyze the speed of all future neurons and the level of control when we attack\footnote{The details on the construction of the neuron wiggle for the attack will be discussed in the following  \autoref{sec:sign_recovery-algorithm}.} the particular neuron across several dual points. 
%At each dual point and for each future neuron the speed is computed as $s_{\textrm on, off} = |v\left(\vec x_{\textrm dual}\right)-v\left(\vec x_{\textrm t}\right)| / \Delta_{\textrm{on,off}}$, i.e., the change in the future neuron's value before its respective ReLU divided by the measured distance $\Delta_{\textrm{on,off}}$ to the first actual future neuron toggling point on the on-, respectively off-side. \\

% ================================================
\subsection{Algorithm}
\label{sec:sign_recovery-algorithm}
% ================================================
%
Let $\xdual$ be a dual point for the target neuron. Assuming momentarily that we do know the correct sign of the weights, we let $\unitvec n$ be the unit normal vector of the ReLU plane pointing towards the direction where the neuron is positive. Walking in the direction of $\unitvec n$ will produce a maximal rate of change in the target neuron, which in turn may toggle neurons in future layers more often. On the other hand, walking in the direction of $-\unitvec n$ will produce no change in the target neuron since it is being suppressed by the ReLU, and may be less likely to toggle neurons in future layers. Of course, this general rule may not hold depending on the unpredictable effect that the displacements have on non-target neurons. However, when conducting many experiments that explore dual points throughout different regions of the input space, we expect that walks on regions where the target neuron is known to be on will on average trigger future neurons faster than those where the target neuron is off.
%
% \begin{algorithm}[htb]
%     \caption{\textsc{RecoverSign}$(i,\mathbf W,\mathbf B)$}
%     \begin{algorithmic}[1]
%         \Require $i$ the index of the target neuron and $\mathbf W,\mathbf B$ the weights and biases of the network up to the target layer (with the target layer known only up to sign per neuron).
%         \Ensure $+1$ if the sign of the target neuron is correct, otherwise $-1$
%         \State $\mathtt{votes}_+\gets 0$
%         \State $\mathtt{votes}_-\gets 0$
%         \For{$\mathtt{experiment} = 1,\ldots,\mathtt N_{exp}$}
%             \State $\mathbf x_d \gets \textsc{RandomDualPoint}(i)$
%             \State $F \gets \textsc{LocalMatrix}(\mathbf W, \mathbf B, \mathbf x_d)$
%             \State $\unitvec n \gets F[:,i]~/~\norm{F[:,i]}$
%             \State $d_+ \gets \textsc{DistanceToToggle}(x_d, \unitvec n)$
%             \State $d_- \gets \textsc{DistanceToToggle}(x_d, -\unitvec n)$
%             \State {\bf if} $d_+ < d_-$ {\bf : } $\mathtt{votes}_+ = \mathtt{votes}_+ +1$
%             \State {\bf else : } $\mathtt{votes}_- =\mathtt{votes}_- + 1$
%         \EndFor
%         \State {\bf if} $\mathtt{votes}_+ > \mathtt{votes}_-$ {\bf : return } $+1$
%         \State {\bf else: return } $-1$
%     \end{algorithmic}
%     \label{alg:sign}
% \end{algorithm}
%
Based on the above argument, we devise Algorithm~\ref{alg:sign}: we assume that we know the correct sign of the neuron, and measure the distance that we need to walk on either side of the ReLU before a future-layer neuron is toggled. These distances are compared, and the experiment is repeated at many different dual points. If our guess for the sign was correct, we expect that a majority of experiments will walk a shorter distance on the on-side than on the off-side. Otherwise, we conclude that the real sign is opposite to our guess.
\begin{algorithm}[htb]
    \footnotesize
    \caption{\footnotesize{\textsc{RecoverSign}$(i,\mathbf W,\mathbf B)$}}
    \begin{algorithmic}[1]
        \Require $i$ the index of the target neuron and $\mathbf W,\mathbf B$ the weights and biases of the network up to and including the target layer (with the target layer known only up to sign per neuron).
        \Ensure $+1$ if the sign of the target neuron is correct, otherwise $-1$
        \State $\mathtt{votes}_+\gets 0$
        \State $\mathtt{votes}_-\gets 0$
        \For{$\mathtt{experiment} = 1,\ldots,\mathtt N_{exp}$}
            \State $\mathbf \xdual \gets \textsc{RandomDualPoint}(i)$
            \State $F \gets \textsc{LocalMatrix}(\mathbf W, \mathbf B, \xdual)$
            \State $\unitvec n \gets F[:,i]~/~\norm{F[:,i]}$
            \State $\DeltaX_+ \gets \textsc{DistanceToToggle}(\xdual, \unitvec n)$
            \State $\DeltaX_- \gets \textsc{DistanceToToggle}(\xdual, -\unitvec n)$
            \State {\bf if} $\DeltaX_+ < \DeltaX_-$ {\bf : } $\mathtt{votes}_+ = \mathtt{votes}_+ +1$
            \State {\bf else : } $\mathtt{votes}_- =\mathtt{votes}_- + 1$
        \EndFor
        \State {\bf if} $\mathtt{votes}_+ > \mathtt{votes}_-$ {\bf : return } $+1$
        \State {\bf else: return } $-1$
    \end{algorithmic}
    \label{alg:sign}
\end{algorithm}
% \end{minipage} % 2C
% \hfill % 2C
% \begin{minipage}[t]{0.48\linewidth} % 2C

% \end{minipage} % 2C
% \end{figure} % 2C
%
\begin{figure}
    \begin{subfigure}{0.0\textwidth}
    \phantomsubcaption\label{fig:distance-a}
    \phantomsubcaption\label{fig:distance-b}
    \phantomsubcaption\label{fig:distance-c}
    \end{subfigure}
    \centering
    \includegraphics[width=1\linewidth]{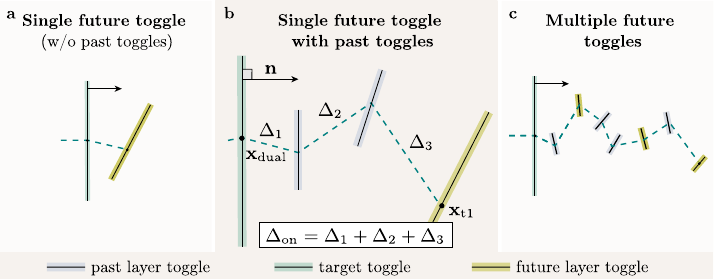}
    \caption{Visualization of the distance measurement using the \DistanceToToggle\xspace algorithm.
    \textbf{a.} In its simplest variant the distance measurement only accepts distance measurements from dual points at which the first toggle is a future layer toggle. \textbf{b.} A slightly more elaborate variant of the distance measurement handles past layer toggles by recomputing the decision hyperplane normal vector and continue moving until the first future layer toggle is encountered. 
    \textbf{c.} In its most elaborate variant the distance measurement can gain statistics at a single dual point, by moving through multiple (instead of only a single) future toggle. }
    \label{fig-distance}
\end{figure}
The precise way in which we perform our walk is visualized in~\autoref{fig-distance} (and detailed in Algorithm~\ref{alg:distance}). In a black box approach we cannot observe the values of future-layer neurons, but we know that the decision plane will change directions whenever any neuron toggles. Therefore, we perform our walk over the decision plane (moving not in the direction of $\pm \unitvec n$ but its projection onto the decision plane), and infer that there has been a neuron toggle whenever we notice a change in the direction of the plane. \\
%% FUTURE TOGGLES ONLY: 
It is important to stress that the toggle should only be counted if it comes from a future layer, since toggles in pasts layers are unaffected by whether the target neuron is on or off. We can easily check if the toggle is from a past layer, since the weights and biases of all past-layer neurons are already known. \\
A unique case of future-neuron influence occurs in the penultimate layer, where no further ReLU layers (and thus no toggles) exist; only the output layer follows. Here, the output logits, though inaccessible in the hard-label scenario, change with the target neuron's output. The current decision boundary eventually intersects with another, allowing us to evaluate this crossing distance, similar to assessing distances to future toggles. \\ 
%%% Handling past or current layer toggles
Once we detected a non-future layer toggle we have two options: Either we decide to discard the dual point altogether (\autoref{fig:distance-a}), or, we try to follow the decision hyperplane \emph{through} the unwanted toggle and continue our search for a future toggle (\autoref{fig:distance-b}). 
In our experiments we choose this middle variant which aims to handle non-future layer toggles, and measures the distance $\DeltaX$ at each input point $\xdual$ by summing the entire walked distance along the decision hyperplane until one future toggle is encountered. \\
The handling of non-future layer toggles does not always succeed, and we still discard the dual point if we keep on encountering the same past layer neurons. Therefore, to successfully perform $\nexp$ experiments we usually  need to investigate a larger number of $\ndual$ dual points.

%\subsubsection{Advanced Distance Measurement Techniques.}
%\TODO{\autoref{fig-distance} somewhere here and explain how we move through past toggles}
%\TODO{We need to explain that an experiment can fail when we get stuck in past layer toggles}, i.e. to successfully perform $\nexp$ experiments we need to investigate a larger number of $\ndual$ dual points. 

%\TODO{We no longer use multiple toggles} Experimental evidence has shown that the results are more reliable when we stop the walk after crossing through a fixed small number of future-layer toggles, rather than always stopping at the first one, so we introduced also the \texttt{nToggles} parameter to control this. \commentAnna{Let us remove this sentence} There are multiple ways in which the ``total'' distance could be measured, but we found it most reliable to measure the distance from the the final point in the walk to the closest point in the initial target neuron's ReLU plane \commentAnna{(no longer true, we use the sum over all distances)}.

%\FloatBarrier
% ================================================
\subsubsection{Confidence Level}\label{sec:confidence-level}%
% ================================================
%%%%% explain p_observed
In our experiment, after conducting $\nexp$ trials (or, in other words, successfully analyzing $\nexp$ dual points), we calculate the observed probability for one sign decision (either + or -) as $p_{\textrm{exp}}=\frac{\max(\votes_+,\votes_-)}{\nexp}.$
% \begin{equation}
% p_{\textrm{exp}}=\frac{\max(\votes_+,\votes_-)}{\nexp}.
% \end{equation}
While the average future neuron speed $s_{\textrm{on}}$ is typically higher on the on-side, this is not consistent across every single dual point (refer to \autoref{tab:intuition}). Thus, we might observe a vote for the correct side with a probability such as $p_{\textrm{exp}}=0.55$. 
%%%%% explain how to distinguish p_observed vs p0+delta
To ensure our observed probability $p_{\textrm{exp}}$ is not simply a result of random variation around $p_0=0.5$, we assess the probability of error, that is the significance level $\alpha$. The significance level is connected to the confidence level $\CL$ by $\CL = 100\times (1-\alpha)\%$.  
% \begin{equation}
% \CL = 100\times (1-\alpha)\%.
% \end{equation}
Hoeffding's inequality~\cite{hoeffding1994probability} provides a bound on the error probability $\alpha$, stating: 
$\alpha = P\left(p_{\textrm{exp}} - p_0>\delta_{\textrm{p}}\right) \leq \exp{\left(-2\delta_{\textrm{p}}^2 \nexp\right)}$. 
% \iffalse
% \begin{equation}
% \alpha = P\left(p_{\textrm{exp}} - p_0>\delta_{\textrm{p}}\right) \leq \exp{\left(-2\delta_{\textrm{p}}^2 \nexp\right)}. 
% \end{equation}
% \fi
% \begin{equation}
% \alpha = P\left(p_{\textrm{exp}} - p_0>\delta_{\textrm{p}}\right) \leq e^{-2\delta_{\textrm{p}}^2 \nexp}
% \end{equation}
%
This implies that the number of trials $\nexp$ required to achieve a specified error probability $\alpha$ is $\nexp \geq \ln(1/\alpha)/(2\delta_\textrm{p}^2)$. As a practical example, at least $\nexp \gtrapprox 1000$ trials are required to achieve a confidence level of $\CL=99\%$ for $p_\textrm{exp}=0.55$. 
If we don't find $p_{\textrm{exp}}$ to be close to $p_0=0.5$, a much smaller minimum number of trials, such as $\nexp \gtrapprox 100$ (for $p_{\textrm{exp}}=0.65$) or $\nexp \gtrapprox 10$ (for $p_{\textrm{exp}}=1.0$).

%\FloatBarrier
% ================================================
\section{Experiments}
\label{sec:experiments}
% ================================================

In our attack on \nameref{sec:our-network} we first discuss the signature recovery (\autoref{sec:experiments-signature}), and then the sign recovery (\autoref{sec:experiments-sign}). In practice, the recovery process begins with signature recovery for hidden layer~1, followed by sign recovery. Once this layer is fully recovered, the process is repeated sequentially for each subsequent layer until the entire network is reconstructed. In all our experiments the current target layer is attacked using standard 64~bit floating point precision, we do, however, assume that the prior layers were perfectly extracted.

\ifePrint 
    Our software implementation will be made available soon in
    \begin{center}
        \texttt{https://github.com/X}
    \end{center}
\else 
    Our software implementation is available in \footnote{Anonymized for submission.}
    \begin{center}
        \texttt{https://github.com/X}
    \end{center}
\fi

% ================================================
\subsubsection{Our CIFAR10 Network}\label{sec:our-network}%
% ================================================
CIFAR-10 is a widely used benchmark dataset in the field of visual deep learning. It is a balanced subset of the larger 80 Million Tiny Images dataset, containing ten classes, including categories like airplanes, cats, and frogs \cite{torralba200880}. Each class consists of images with RGB channels of size 
$32\times32$ pixels, resulting in 3,072 pixels per image. The dataset comprises a total of 50,000 training images and 10,000 test images.

Our model for CIFAR-10 employs 3,072 input neurons (one per pixel) and consists of four densely connected hidden layers. The first three hidden layers contain 256 neurons each, while the fourth layer contains 64 neurons, all using ReLU activations. The network's output layer has 10 neurons with softmax activation. 
We follow the same training procedure\footnote{For training, we use standard preprocessing and optimization techniques. The pixel values are rescaled from the original range of 0 to 255 to a range of 0 to 1. The model is trained using stochastic gradient descent (SGD) with a momentum of 0.9 and sparse categorical cross-entropy as the loss function. A batch size of 64 is used during training.} as~\cite{lin2015far,canales2023polynomial}.

Our four-hidden-layer model achieves a test accuracy of 0.52 on CIFAR-10, which aligns with the expected performance for densely connected neural networks utilizing pure ReLU activation functions, typically around 0.53~\cite{lin2015far}. It is important to note that higher test accuracies are attainable with more sophisticated neural network architectures. For example, Google Research's Vision Transformer (ViT-H/14) currently achieves a state-of-the-art test accuracy of 0.995 on CIFAR-10 \cite[Table 2]{dosovitskiy2020image}.

% ================================================
\subsection{Signature Recovery Attack}
\label{sec:experiments-signature}
% ================================================

We have now each of the major algorithms necessary to recover the signatures of the model.

\subsubsection{Description of the Implementation.} 
The total runtime for the signature recoveries of each target layer can be expressed as 
\begin{equation}
t_{\textrm{signatures}} = t_{\textrm{dual}} + t_{\textrm{cluster}} + t_{\textrm{unify}}, 
\end{equation}
where 	
$t_{\textrm{dual}}$ is the time taken to find dual points, 
$t_{\textrm{cluster}}$ is the time taken to cluster the dual points (cf. step~1 in \autoref{sec:deeper-layer-signature}), i.e. identify which target neuron in the target layer they belong to, and,
$t_{\textrm{unify}}$ the time taken to unify the corresponding dual spaces (cf. step~2 in \autoref{sec:deeper-layer-signature}). 

Our proof-of-concept implementation disregards the runtime of the $n^2$ pairwise clustering $t_{\textrm{cluster}}$ needed to identify dual points that are mutually consistent. Although this method is known to work in principle (we validated it post-hoc) and clearly runs in polynomial time, applying it naively (i.e., without many potential optimizations) to a real attack would involve a $(1 \text{ million})^2$ time complexity, requiring weeks of computation. \\
%I improve efficiency by, at decision boundary points, computing the gradient symbolically instead of with calls to binary search. This improves performance by a constant factor.
Further, efficiency is improved by computing the gradient symbolically at decision boundary points, instead of relying on binary search, thereby enhancing performance by a constant factor, and achieving a shorter $t_{\textrm{dual}}$.

\subsubsection{Consistency Verification}

Recall that, after generating dual points, the first step in our attack is to
cluster dual points according to whether or not they are \emph{consistent}.
To implement this in a numerically stable manner, 
we collect a set of inputs that form a basis to each of the two partial dual spaces,
and then compute the singular value decomposition on this set of points.
By inspecting the smallest singular value, we can see if the dual spaces are consistent,
because consistent dual spaces will not span $\mathbb{R}^{d_0}$ whereas inconsistent
ones will.
As we can see in Figure~\ref{fig:consistent}, running the algorithm on dual points that
come from the same neuron result in a significantly different distribution of singular
values than when the algorithm is run on dual points from different neurons.

% \begin{figure}
% \centering
%     \includegraphics[scale=.7]{images/consistent.pdf}
%     \caption{Our method that computes the consistency between two dual points
%     nearly perfectly separates the distribution of consistent and inconsistent neurons.}
%     \label{fig:consistent}
% \end{figure}

In a very small number of cases, 
we find that different neurons can incorrectly have very low singular values.
To address this rare occurrence, we apply a secondary filter where we consider
triples of dual points $(a,b,c)$ that are believed to be consistent with each other;
if the triple is inconsistent, we discard the all three.
Empirically we observe that this completely eliminates all false positives.

\subsubsection{Identifying a diverse set of dual points}

% \begin{figure}
%     \centering
%     \includegraphics[scale=.9]{images/numqueries.pdf}
%     \caption{Fraction of inputs on each layer recovered as a function of the
%     number of dual points explored. 
%     Early layers are easier to recover than later layers, requiring just a few 
%     thousand dual points, but later layers can require millions of dual points.}
%     \label{fig:numqueries}
% \end{figure}

In order for our algorithm to recover neurons at deeper layers,
we must collect a \emph{diverse} set of inputs that cause every input to the
neuron to be active at least once.
Unfortunately, we find that some neurons in the model we have trained are \emph{almost dead},
and rarely activate.
As a result, it is necessary to identify a very large number of dual points in order
to reconstruct these last few parameters.

Figure~\ref{fig:numqueries} summarizes this analysis.
Each layer requires more queries to extract than the previous,
because

\subsubsection{Measuring Extraction Fidelity}

% \begin{figure}
%     \centering
%     \includegraphics[scale=.6]{images/bit_precision.pdf}
%     \caption{}
%     \label{fig:fidelity}
% \end{figure}

As mentioned earlier, throughout this paper we describe an algorithm that
succeeds as long as arithmetic is performed to an arbitrary level of precision.
However, in our implementation we make use of 64-bit floating point arithmetic.
In this section we validate that our attack correctly recovers the weights of
the model up to a high degree of precision, assuming that all prior layers have
been perfectly recovered.
As we can see in Figure~\ref{fig:fidelity}, all neurons are extracted with extremely
high fidelity---usually 18 or more bits of precision.

This proof-of-concept assessment of the signature-recovery algorithm (neglecting $t_{\textrm{cluster}}$, and with improved $t_\nabla$)  had been successfully implemented, and has a running time of approximately 16~hours on a 256-core machine with additional GPU support.

% ================================================
\subsection{Sign Recovery Attack}
\label{sec:experiments-sign}
% ================================================

We will first describe assumptions and possible further optimizations of our current implementation of the sign recovery attack, and then conclude with the obtained results. 

% ================================================
\subsubsection{Description of the Implementation}%
% ================================================
The total runtime for the sign recovery of each target neuron can be expressed as: 
\begin{equation}
t_\textrm{sign} = \left( t_\textrm{dual} + t_{m} + t_\textrm{vote} \right) \times \ndual, 
\end{equation}
where $t_\textrm{dual}$ is the time needed to identify a single dual point for the target neuron, 
$t_{m}$ is the time needed to determine the decision hyperplane normal vectors ($\vec m_{\textrm{on}}$, $\vec m_{\textrm{off}}$), 
$t_\textrm{vote}$ is the actual time needed to execute the attack on the target neuron and obtain a single vote for sign recovery, 
and $\ndual$ is the total number of dual points required to attain the desired confidence level~(cf.~\autoref{sec:confidence-level}) for the attacked neuron.  \\
In our current implementation, we assume the dual points and decision hyperplane vectors are predetermined, as these components are common to both sign recovery and signature recovery. Notably, here we use as inputs critical points for the target neuron derived from a set of random uniformly distributed points over the interval $[-5,+5]^{d_0}$.
\\
During the sign recovery process for a single neuron, four CPU cores are utilized. On a 256-core computer, this allows for the simultaneous recovery of 64 neurons. A potential further  optimization is parallelized recovery on the dual point level, as multiple dual points can, in principle, independently be analyzed in parallel. Each vote evaluation is independent, enabling simultaneous processing. \\
Another potential optimization involves adaptive control of the number of dual points investigated. Currently, we use a fixed minimum number of input points ($\nexp^\textrm{min}=100$ for hidden layers 1 and 4, and $\nexp^\textrm{min}=1,000$ for hidden layers 2 and 3 as detailed in \autoref{sec:confidence-level}). This fixed minimum is often higher than necessary for many of the targeted neurons.

% ================================================
\subsubsection{Results}%
% ================================================
We use the confidence level $\CL$ introduced in \autoref{sec:confidence-level} to monitor the sign recovery. \autoref{fig-sign-recovery-results} shows examples for the evolution of $\CL$ with the number of investigated dual points $\ndual$ in hidden layers 1, ..., 4. 
%%%%%%%%%%%%%%%%%%%%%%%%%%%%%%%%%%%%%%%%%%%%%
% This figure has been generated using the code in 
% /home/anna/file-system/pure/share/users/anna_pub/2024-09_DETI-II/D/03b_CreateResultsFigure.ipynb
% \begin{figure}[!htb]
%     \centering
%     \includegraphics[width=1\linewidth]{images/fig-sign-recovery-results.png}
%     \caption{The evolution of the sign recovery confidence level $\CL$ with the investigated number of dual points $\ndual$ for easy neurons, average neurons, and the most difficult neurons in each layer.}
%     \label{fig-sign-recovery-results}
% \end{figure}
In hidden layer 1 the likelihood of encountering current layer toggles is very low, allowing for a successful investigation of every dual point. Further, we have perfect control over the target layer, i.e. $p_{\textrm{exp}}=100\%$. Consequently, after analyzing approximately $\nexp=\ndual\approx 10$ points, we achieve a confidence level of $\CL\approx 100\%$ for all neurons. \\ 
Similarly, in hidden layer 4, we expect strong control over the target layer. However, this layer is characterized by a high likelihood of encountering current or past layer toggles, which results in variability in the number of dual points $\ndual \approx 876$ required to attain the final confidence level (see~\autoref{tab:sign-recovery-results} for more details). \\ 
In hidden layers 2 and 3, the behavior of the confidence level with respect to the number of dual points shows greater variability. While the easiest neurons achieve confidence levels of around 100\% after approximately $\ndual\approx 250$,  average neurons require about $\ndual \approx 750$ dual points to reach similarly high confidence levels~(cf.~\autoref{fig-sign-recovery-results}). Notably, there are six and seven particularly challenging neurons in layers 2 and 3, respectively, that maintain $CL\leq 75\%$ even after extensive analysis. For these neurons we re-run the sign recovery on a fresh set of input points and can indeed achieve higher confidence levels. 

%\vspace*{-\baselineskip}
%%%%%%%%%%%%%%%%%%%%%%%%%%%%%%%%%%%%%%%%%%%%%%
% The data in this table has been generated using 
% /home/anna/file-system/pure/share/users/anna_pub/2024-09_DETI-II/D/run_exp_create_report.ipynb
%\setlength{\tabcolsep}{6pt}
\begin{table}[htb]
\vspace*{-2\baselineskip}
\begin{threeparttable}
\caption{Summary of sign recovery results on our CIFAR10 network.}
\label{tab:sign-recovery-results}
\begin{tabular}{@{}p{\textwidth}@{}}
\centering
    %%%%% tabular part %%%%%%%%%
    \resizebox{\textwidth}{!}{%
    \footnotesize{
    \begin{tabular}{ll c ccc cc}
    \toprule
        &  
        &
        & \multicolumn{3}{c}{\textbf{Confidence level}}
        & \multicolumn{2}{c}{\textbf{Times per neuron}}
        \\
        \textbf{Layer}
        & \textbf{Recovered}
        & \textbf{$\ndual$}
        & \small{$\min(\CL)$} 
        & \small{$\textrm{mean}(\CL)$} 
        & \small{$\max(\CL)$} 
        & $\overline{t}_{\textrm{vote}}$ 
        &  $\sum t_{\textrm{vote}}$ 
        \\ 
        \cmidrule(rl){1-2} \cmidrule(lr){3-3} \cmidrule{4-6} 
        \cmidrule(lr){7-8}  
        1
        & \textbf{256}/256
        & (100$\pm$0)\,s 
        & 100\% & 100\%  & 100\% 
        & ($0.71\pm 0.13)$\,s 
        & ($220\pm 127)$\,s\\
        2
        & \textbf{256}/256\tnote{*}
        & ($1142\pm268$)
        & 86.27\% & 99.39\%  & 100.00\%
        & ($0.29\pm0.07$)\,s
        & (1670$\pm$1131)\,s\\
        3
        & \textbf{256}/256\tnote{*}
        & (1275$\pm$193)
        & 88.06\%& 99.48\% & 100.00\%
        & (2.16$\pm$0.43)\,s
        & (3468$\pm$485)\,s\\
        4
        & \textbf{64}/64
        & (876$\pm$377)
        & 98.17\% & 99.63\% & 99.69\%
        & (1.48$\pm$0.39)\,s
        & (2311$\pm$661)\,s\\
    \bottomrule 
    \end{tabular}%
    }}
    %%%%%%%%%%%%%%%%%%%%%%%%%%%%
    \end{tabular}
    \begin{tablenotes}
    \item[] \footnotesize{$\ndual$: The average of the dual points across all neurons.
    $\overline{t}_{\textrm{vote}}$: The average of the runtime per dual point. 
    $\sum t_{\textrm{vote}}$: The average of the total runtime per neuron. 
    \item[*] In layer 2 and 3, six, respectively seven neurons finalized the first run with $\CL<75\%$ and were re-run a second time to confirm the sign votes and achieve higher confidence levels.} 
    \end{tablenotes}
\end{threeparttable}
\vspace*{-\baselineskip}
\end{table}

The sign recovery process for all neurons across all layers yields correct results. 
\autoref{tab:sign-recovery-results} summarizes the sign recovery results.
This proof-of-concept assessment of the sign-recovery algorithm indicates that given the indicated number of input points and their decision hyperplane normal vectors, i.e. $t_\textrm{sign} \approx  t_\textrm{vote}$, we can run our neuron sign recovery attack on 64 neurons in parallel on our 256 core server and recover hidden layers $1,\ldots,4$ in around 8.5\,hours runtime.\footnote{
This estimate is based on \autoref{tab:sign-recovery-results}: 
In detail we can recover the signs of 
% 4*220/60 = 14.6667 
hidden layer~1 in four batches with a total runtime of about $4\times220\,$s$\approx 15$ minutes, 
% (1670*4+1670)/(60*60) = 2.3194  
hidden layer~2 in about $(4+1)\times1,670\,$s$\approx$2.5\,hours (note that we added one extra-run for the initially low-confidence neurons.), 
% (3468*4+3468)/(60*60) = 4.8167  
hidden layer~3 in about 5\,hours, 
% 2311/60 = 38.5167 
and hidden layer 4 in 40\,minutes.}

% ================================================
\section{Conclusions}
\label{sec:conclusions}
% ================================================
In this paper we have finally solved the long standing open problem of how difficult it is to extract all the secret parameters of a ReLU-based DNN by interacting with its black box implementation. While previous papers have shown that this problem can be solved in polynomial time in the easiest attack scenario (in which the attacker is given the precise numeric values of all the DNN's logits), in this paper we develop the first polynomial time attack even in the hardest attack scenario (in which only the hard-labels of chosen inputs are provided). 
In our proof-of-concept implementation we demonstrated the correctness of all the elements of our attack by successfully extracting all the approximately one million parameters of a realistic CIFAR10 network with about a thousand neurons. However, a fully optimized end-to-end blackbox implementation which can be used by third parties is left to future work.
% \proposeRemove{We demonstrated the success of our attack by extracting all the approximately one million parameters of a realistic CIFAR10 network with about a thousand neurons. }

\ifePrint
    \begin{credits}
    \subsubsection{\ackname}
We would like to express our sincere gratitude to Isaac Canales-Martínez for his valuable contributions to this paper. His experimental tests provided critical insights that shaped our findings. We also appreciate his  proofreading of the manuscript, which enhanced its clarity and coherence. Additionally, his review of the code written by others helped to maintain the integrity and accuracy of our work.     
    \subsubsection{\discintname}
    The authors have no competing interests to declare that are relevant to the content of this article.
    \end{credits}
\else
\begin{credits}
%\subsubsection{\ackname}
%A bold run-in heading in small font size at the end of the paper is used for general acknowledgments, for example: This study was funded by X (grant number Y).

\subsubsection{\discintname}
The authors have no competing interests to declare that are relevant to the content of this article.
\end{credits}
\fi

\bibliographystyle{splncs04}
\bibliography{bibliography}

\begin{thebibliography}{10}
\providecommand{\url}[1]{\texttt{#1}}
\providecommand{\urlprefix}{URL }
\providecommand{\doi}[1]{https://doi.org/#1}

\bibitem{Baum90a}
Baum, E.B.: A polynomial time algorithm that learns two hidden unit nets.
  Neural Comput.  \textbf{2}(4),  510--522 (1990),
  \url{https://doi.org/10.1162/neco.1990.2.4.510}

\bibitem{Baum91}
Baum, E.B.: Neural net algorithms that learn in polynomial time from examples
  and queries. {IEEE} Trans. Neural Networks  \textbf{2}(1),  5--19 (1991),
  \url{https://doi.org/10.1109/72.80287}

\bibitem{BlumR93}
Blum, A., Rivest, R.L.: Training a 3-node neural network is {NP-Complete}. In:
  Hanson, S.J., Remmele, W., Rivest, R.L. (eds.) Machine Learning: From Theory
  to Applications - Cooperative Research at Siemens and {MIT}. Lecture Notes in
  Computer Science, vol.~661, pp. 9--28. Springer (1993)

\bibitem{canales2023polynomial}
Canales-Mart{\'\i}nez, I.A., Chavez-Saab, J., Hambitzer, A.,
  Rodr{\'\i}guez-Henr{\'\i}quez, F., Satpute, N., Shamir, A.: Polynomial time
  cryptanalytic extraction of neural network models. Cryptology ePrint Archive
  (2023)

\bibitem{carlini2020cryptanalytic}
Carlini, N., Jagielski, M., Mironov, I.: Cryptanalytic extraction of neural
  network models. In: Annual International Cryptology Conference. pp. 189--218.
  Springer (2020)

\bibitem{chen2024}
Chen, Y., Dong, X., Guo, J., Shen, Y., Wang, A., Wang, X.: Hard-label
  cryptanalytic extraction of neural network models. Cryptology {ePrint}
  Archive, Paper 2024/1403 (2024), \url{https://eprint.iacr.org/2024/1403}

\bibitem{abs-2105}
Daniely, A., Granot, E.: An exact poly-time membership-queries algorithm for
  extraction a three-layer relu network. CoRR  \textbf{abs/2105.09673} (2021),
  \url{https://arxiv.org/abs/2105.09673}

\bibitem{dosovitskiy2020image}
Dosovitskiy, A., Beyer, L., Kolesnikov, A., Weissenborn, D., Zhai, X.,
  Unterthiner, T., Dehghani, M., Minderer, M., Heigold, G., Gelly, S., et~al.:
  An image is worth 16x16 words: Transformers for image recognition at scale.
  arXiv preprint arXiv:2010.11929  (2020)

\bibitem{Fefferman1994}
Fefferman, C.: Reconstructing a neural net from its output. Revista Matemática
  Iberoamericana  \textbf{10}(3),  507--555 (1994),
  \url{http://eudml.org/doc/39464}

\bibitem{Foerster24}
Foerster, H., Mullins, R.D., Shumailov, I., Hayes, J.: Beyond slow signs in
  high-fidelity model extraction. CoRR  \textbf{abs/2406.10011} (2024).
  \doi{10.48550/ARXIV.2406.10011},
  \url{https://doi.org/10.48550/arXiv.2406.10011}

\bibitem{HancockGM94}
Hancock, T.R., Golea, M., Marchand, M.: Learning nonoverlapping perceptron
  networks from examples and membership queries. Mach. Learn.  \textbf{16}(3),
  161--183 (1994). \doi{10.1007/BF00993305},
  \url{https://doi.org/10.1007/BF00993305}

\bibitem{hoeffding1994probability}
Hoeffding, W.: Probability inequalities for sums of bounded random variables.
  The collected works of Wassily Hoeffding pp. 409--426 (1994)

\bibitem{jagielski2020high}
Jagielski, M., Carlini, N., Berthelot, D., Kurakin, A., Papernot, N.: High
  accuracy and high fidelity extraction of neural networks. In: 29th USENIX
  security symposium (USENIX Security 20). pp. 1345--1362 (2020)

\bibitem{lin2015far}
Lin, Z., Memisevic, R., Konda, K.: How far can we go without convolution:
  Improving fully-connected networks. arXiv preprint arXiv:1511.02580  (2015)

\bibitem{MartinelliSGB24}
Martinelli, F., Simsek, B., Gerstner, W., Brea, J.: Expand-and-cluster:
  Parameter recovery of neural networks. In: Forty-first International
  Conference on Machine Learning, {ICML} 2024, Vienna, Austria, July 21-27,
  2024. OpenReview.net (2024)

\bibitem{MilliSDH19}
Milli, S., Schmidt, L., Dragan, A.D., Hardt, M.: Model reconstruction from
  model explanations. In: danah boyd, Morgenstern, J.H. (eds.) Proceedings of
  the Conference on Fairness, Accountability, and Transparency, FAT* 2019,
  Atlanta, GA, USA, January 29-31, 2019. pp.~1--9. {ACM} (2019)

\bibitem{Reith0T19}
Reith, R.N., Schneider, T., Tkachenko, O.: Efficiently stealing your machine
  learning models. In: Cavallaro, L., Kinder, J., Domingo{-}Ferrer, J. (eds.)
  Proceedings of the 18th {ACM} Workshop on Privacy in the Electronic Society,
  WPES@CCS 2019, London, UK, November 11, 2019. pp. 198--210. {ACM} (2019)

\bibitem{RolnickK20}
Rolnick, D., K\"ording, K.P.: Reverse-engineering deep {ReLU} networks. In:
  Proceedings of the 37th International Conference on Machine Learning, {ICML}
  2020, 13-18 July 2020, Virtual Event. Proceedings of Machine Learning
  Research, vol.~119, pp. 8178--8187. {PMLR} (2020)

\bibitem{torralba200880}
Torralba, A., Fergus, R., Freeman, W.T.: 80 million tiny images: A large data
  set for nonparametric object and scene recognition. IEEE transactions on
  pattern analysis and machine intelligence  \textbf{30}(11),  1958--1970
  (2008)

\bibitem{tramer2016stealing}
Tram{\`e}r, F., Zhang, F., Juels, A., Reiter, M.K., Ristenpart, T.: Stealing
  machine learning models via prediction $\{$APIs$\}$. In: 25th USENIX security
  symposium (USENIX Security 16). pp. 601--618 (2016)

\end{thebibliography}

\appendix

% ================================================
\section{Appendices}
\label{sec:appendix}
% ================================================

\FloatBarrier
% ================================================
%\section{Preliminaries}
%\label{sec:preliminaries}
% ================================================

\subsection{Basic Definitions and Notation}\label{sec:definitions}

%ADD HERE THE NOTATIONS AND THE ASSUMPTIONS (SUCH AS A FULLY CONNECTED DNN AND SUFFICIENTLY HIGH ARITHMETIC PRECISION) AND MENTION THAT OUR ATTACK CAN FAIL IF WE ARE EXTREMELY UNLUCKY (EG IF A SYSTEM OF LINEAR EQUATIONS IS UNSOLVABLE)
% We should have definitions for

% \begin{definition}
% k-Deep Neural Network [CJM'20]
% \end{definition}

% \begin{definition}
% decision boundary $\mathcal{D}(f_\theta)$ is the set of inputs $x$.
% \end{definition}

% \begin{definition}
% Definite how each layer is a linear transformation [CJM'20]
% \end{definition}

% \begin{definition}
% Neurons $\eta_i$ [CJM'20]
% \end{definition}

% \begin{definition}
% The parameters $\theta$ of a model TODO [CJM'20]
% \end{definition}

Here we present some basic definitions and notations used throughout the manuscript which closely follows the terminology first presented in~\cite{carlini2020cryptanalytic} and then adopted in more recent attacks~\cite{canales2023polynomial,chen2024}.

\begin{definition}%[\cite{carlini2020cryptanalytic}]
\label{def:DNN}
An \emph{$r$-deep neural network} $f_{\theta}$ is a function parameterized by $\theta$ that takes inputs from an input space $\mathcal{X}$ and returns values in an output space $\mathcal{Y}$. The function $f$ is composed as a sequence of functions alternating between linear layers $f_i$ (of different dimensions $d_i$) and a nonlinear function (which acts component-wise) $\sigma$:
\[ f = f_{r+1} \circ \sigma \circ \cdots \circ \sigma \circ f_{2} \circ \sigma \circ f_{1}. \]
\end{definition}

As in \cite{carlini2020cryptanalytic,canales2023polynomial,chen2024}, we study deep neural networks (DNNs) where $\mathcal{X} = \mathbb{R}^{d_0}$, $\mathcal{Y} = \mathbb{R}^{d_{r+1}}$ and $d_0$, \ldots, $d_{r+1}$ are positive integers. Also, we only consider neural networks using the ReLU activation function defined as $\sigma: x \mapsto \max(x, 0)$.

\begin{definition}\label{def:fi-Ai-bi}%[\cite{carlini2020cryptanalytic}]
The $i$-th \emph{fully connected layer} of a neural network is a function $f_i: \mathbb{R}^{d_{i-1}} \rightarrow \mathbb{R}^{d_i}$ given by the affine transformation
\[ f_i(x) = A^{(i)}x + b^{(i)}, \]
where $A^{(i)} \in \mathbb{R}^{d_i \times d_{i-1}}$ and $b^{(i)} \in \mathbb{R}^{d_i}$ are, respectively, the \emph{weight matrix} and the \emph{bias vector} of the $i$-th \emph{layer}, and $d_{i-1}, d_i$ are positive integers.
\end{definition}

\begin{definition}\label{def:neuron}
A \emph{neuron} is a function determined by the corresponding weight matrix followed by an activation function. Particularly, the $j$-th neuron of layer $i$ is the function $\eta$ given by
\[ \eta(x) = \sigma(A_{j}^{(i)}x + b_j^{(i)}), \]
where $A_{j}^{(i)}$ and $b_j^{(i)}$ denote, respectively, the $j$-th row of $A^{(i)}$ and the $j$-th coordinate of $b^{(i)}$. An $r$-deep neural network has $N = \sum_{k=1}^{r} d_k$ neurons.
\end{definition}

\begin{definition}
The \emph{architecture} of a fully connected neural network is described by specifying its number of layers along with the dimension $d_i$ (i.e., number of neurons) of each layer $i =1,\cdots, r+1$. We say that $d_0$ is the dimension of the inputs to the neural network and $d_{r+1}$ is the number of outputs of the neural network.
\end{definition}

\begin{definition}
The \emph{parameters} $\theta$ of an $r$-deep neural network $f_{\theta}$ are the concrete assignments to the weights $A^{(i)}$ and biases $b^{(i)}$, $i \in \{1,2,\dots, r+1\}$.
\end{definition}

When working under the hard-label setting, the raw output $f(x)$ is processed before returning the result. We consider the processing employed in \cite{chen2024}, which is presented in the definition below.

\begin{definition}\label{def:hard-label}
Let $f: \mathcal{X} \rightarrow \mathcal{Y}$ be an $r$-deep neural network with $\mathcal{Y} = \mathbb{R}^{d_{r+1}}$. The \emph{hard-label} $z$ on the outputs $f(x)$ is computed as
$\text{arg max}_i f(x)_i$, the coordinate of the maximum of $f(x)$.\footnote{If multiple entries in $f(x)$ have the same maximum value, the hard-label is the smallest coordinate of these equal entries}
%\comment{CHECK: This definition is from AC24, however, I think the value when $d_{k+1} > 1$ should be the index of the coordinate, not the coordinate itself.}
\end{definition}

\begin{definition}
Let $\mathcal{V}(\eta;x)$ denote the value that neuron $\eta$ takes with $x \in \mathcal{X}$ before applying $\sigma$. If $\mathcal{V}(\eta;x) > 0$ then $\eta$ is \emph{active}. Otherwise, the neuron is \emph{inactive}. The state of $\eta$ on input $x$ (i.e., active, or inactive) is denoted by $\mathcal{S}(\eta;x)$.
% The set of all critical points for a neuron $\eta$ is denoted by $\mathcal{W}(\eta)$.
\end{definition}
%Let $x \in \mathbb{R}^{d_0}$ be an input to the DNN s.t. $\mathcal{V}(\eta;x) = 0$ for a neuron $\eta$. Then, we call $x$ a \emph{critical point} for $\eta$.

\begin{definition}
A critical point $x$ satisfies $\mathcal{V}(\eta; x) = 0$ for some neuron $\eta$
the value $\mathcal{V}(\eta; x) = 0$.
\end{definition}

\begin{definition}\label{def:dual-point}
A \emph{dual} point $d$ is a point that is both on the decision boundary (i.e., $z(f(d + \varepsilon)) \ne z(f(d - \varepsilon))$ for $\epsilon>0$) and on some critical hyperplane (i.e., there is some neuron with $\eta(d) = 0$).
\end{definition}

\begin{definition}
Let $x \in \mathcal{X}$. The \emph{linear neighbourhood} of $x$ is the set
\[ \{ u \in \mathcal{X} \mid \mathcal{S}(\eta;x) = \mathcal{S}(\eta;u) \text{\ for all neurons $\eta$ in the DNN\,} \}. \]
\end{definition}

Since the state of all neurons remains the same for all inputs in the same linear neighbourhood, the DNN behaves as a linear map (within that linear neighbourhood). Let $F_{i,j} = \sigma \circ f_{j} \circ \cdots \circ \sigma \circ f_{i}$, $1 \leq i \leq j \leq r$. Then, for all $x$ in a linear neighbourhood,
\begin{align*}
    F_{i,j}(x) &= I^{(j)}(A^{(j)} \cdots (I^{(i+1)}(A^{(i+1)}(I^{(i)}(A^{(i)}x + b^{(i)})) + b^{(i+1)}) \cdots + b^{(j)}) \\
                &= \Gamma x + \beta,
\end{align*}
where $I^{(\ell)}$ are $0-1$ diagonal matrices with a $0$ on the diagonal's $k$-th entry when neuron $k$ at layer $\ell$ is inactive and $1$ when that neuron is active.

Let $i \in \{1, \dots, r+1\}$. The DNN can be regarded as a composition of an {\it input function} $F_{i-1}$, the $i$-th layer and an {\it output function} $G_{i+1}$:
\[ f = \underbrace{f_{r+1} \circ \sigma \circ \cdots \circ \sigma \circ f_{i+1}}_{G_{i+1}} \circ \,\,\sigma \circ f_i \circ \underbrace{\sigma \circ f_{i-1} \circ \cdots \circ \sigma \circ f_{1}}_{F_{i-1}}. \]
When $i=1$, $F_{0}$ is the identity map on the input; when $i=r+1$, $G_{r+2}$ is the identity map on the output. Furthermore, if we restrict inputs $x'$ to be in the linear neighbourhood of $x \in \mathcal{X}$, we can write $F_{i-1}$ and $G_{i+1}$, respectively, as
\[ F_{i-1}(x') = F^{(i-1)}_{x}x' + b_{x}^{(i-1)} \quad \text{and} \quad  G_{i+1}(x') = G^{(i+1)}_{x}x' + b_{x}^{(i+1)}. \]
% We use the subscript in the collapsed matrices and bias vectors to indicate that they are defined in the linear neighborhood of $x$.
In the context of our attack, $i$ is the index of the target layer and $F_{i-1}, G_{i+1}$ represent, respectively, the recovered and non-recovered layers of the network. This view of the DNN is visually depicted as in \autoref{fig:Network}.

% \begin{figure}[!ht]
%     \centering
%     \includegraphics[width=1\linewidth]{images/fig-nn.png}
%     \caption{Representation of the DNN as a composition of the recovered input function $F_{i-1}$, layer $i$ and an non-recovered output function $G_{i+1}$.}
%     \label{fig:Network}
% \end{figure}

% ------------------------------------------------

\subsection{Algorithms}

\begin{algorithm}[htb]
    \footnotesize
    \caption{\footnotesize{\DistanceToToggle$(\xdual,\unitvec{n})$}}
    \begin{algorithmic}[1]
        \Require A dual point $\xdual$ for the target neuron, %an integer $N$ 
        and $\unitvec{n}$ the normal unit vector of the neuron's ReLU plane. Also uses parameters \texttt{nToggles} with default value 1, and $\delta$ (a small number).
        \Ensure Walks in the direction of $\unitvec n$ projected to the decision plane until $\mathtt{nToggles}$ neurons from future layers are toggled, and returns the total distance walked along the decision hyperplane in the input space.
        \State $\DeltaX \gets 0$
        \State $\vec x_0 \gets \xdual$
        \State $\mathbf{dx} \gets \unitvec n$
        \State \texttt{futureLayerToggles}$\gets 0$
        \While{$\mathtt{futureLayerToggles} < \mathtt{nToggles}$}
            \State $\unitvec m\gets \textsc{DecisionPlaneUnitVector}(\mathbf x_0 + \delta \cdot \mathbf{dx})$
            \State $\mathbf{dx} \gets \unitvec n - \dotp{\unitvec n}{\unitvec m} \unitvec m$ \Comment{Project $\unitvec n$ onto the decision plane}
            \State Walk from $\vec x_0$ in the direction $\mathbf{dx}$ and let $\vec x_1$ be the point at which the 
            \State ... direction of the decision boundary first changes.
            \If{no past- or current-layer neuron being toggled between $\vec x_0$ and $\vec x_1$} 
                \State $\mathtt{futureLayerToggles}\gets\mathtt{futureLayerToggles}+1$
            \Else 
                \State either discard the dual point or handle the past- or current-layer toggle
            \EndIf 
            \State $\DeltaX \gets \DeltaX + ||\vec x_1 - \vec x_d||$ 
            \State $\mathbf x_0 \gets \mathbf x_1$
        \EndWhile
        \Return $\DeltaX$ % $\dotp{\vec x_1 - \vec x_d}{\unitvec n}$
    \end{algorithmic}
    \label{alg:distance}
\end{algorithm}

\subsection{Supporting Figures on Experimental Outcomes}

\begin{figure}
\centering
    \includegraphics[scale=.7]{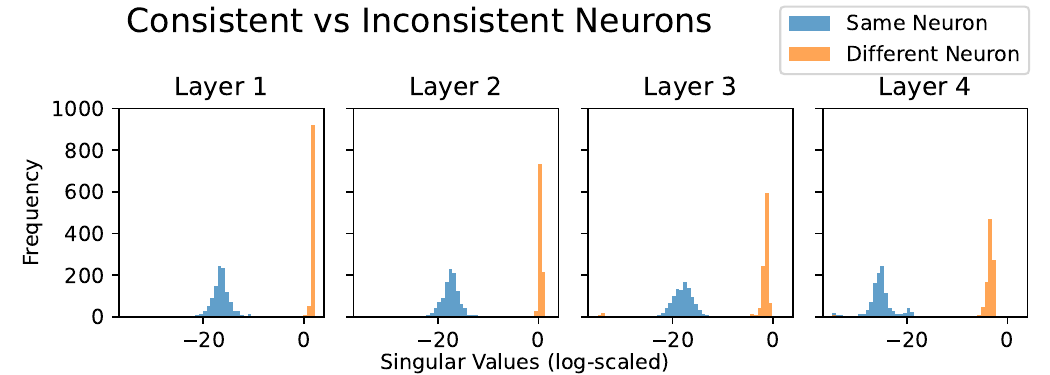}
    \caption{Our method that computes the consistency between two dual points
    nearly perfectly separates the distribution of consistent and inconsistent neurons.}
    \label{fig:consistent}
\end{figure}

\begin{figure}
    \centering
    \includegraphics[scale=.9]{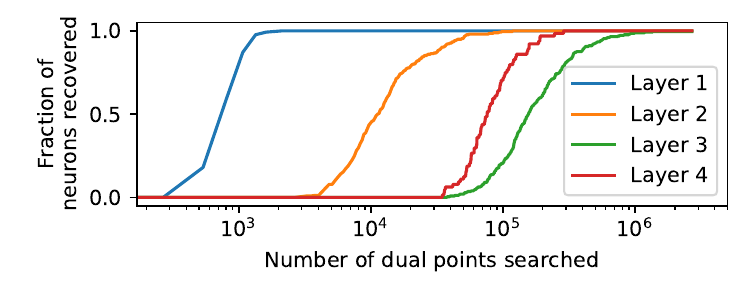}
    \caption{Fraction of inputs on each layer recovered as a function of the
    number of dual points explored. 
    Early layers are easier to recover than later layers, requiring just a few 
    thousand dual points, but later layers can require millions of dual points.}
    \label{fig:numqueries}
\end{figure}

\begin{figure}
    \centering
    \includegraphics[scale=.6]{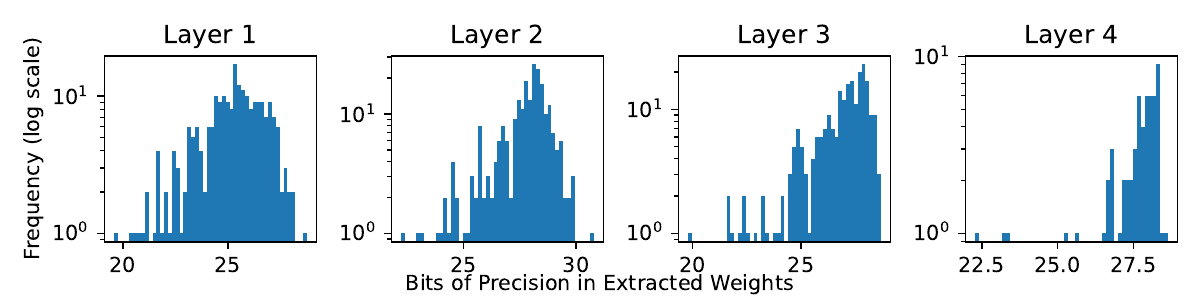}
    \caption{}
    \label{fig:fidelity}
\end{figure}

\begin{figure}[!htb]
    \centering
    \includegraphics[width=1\linewidth]{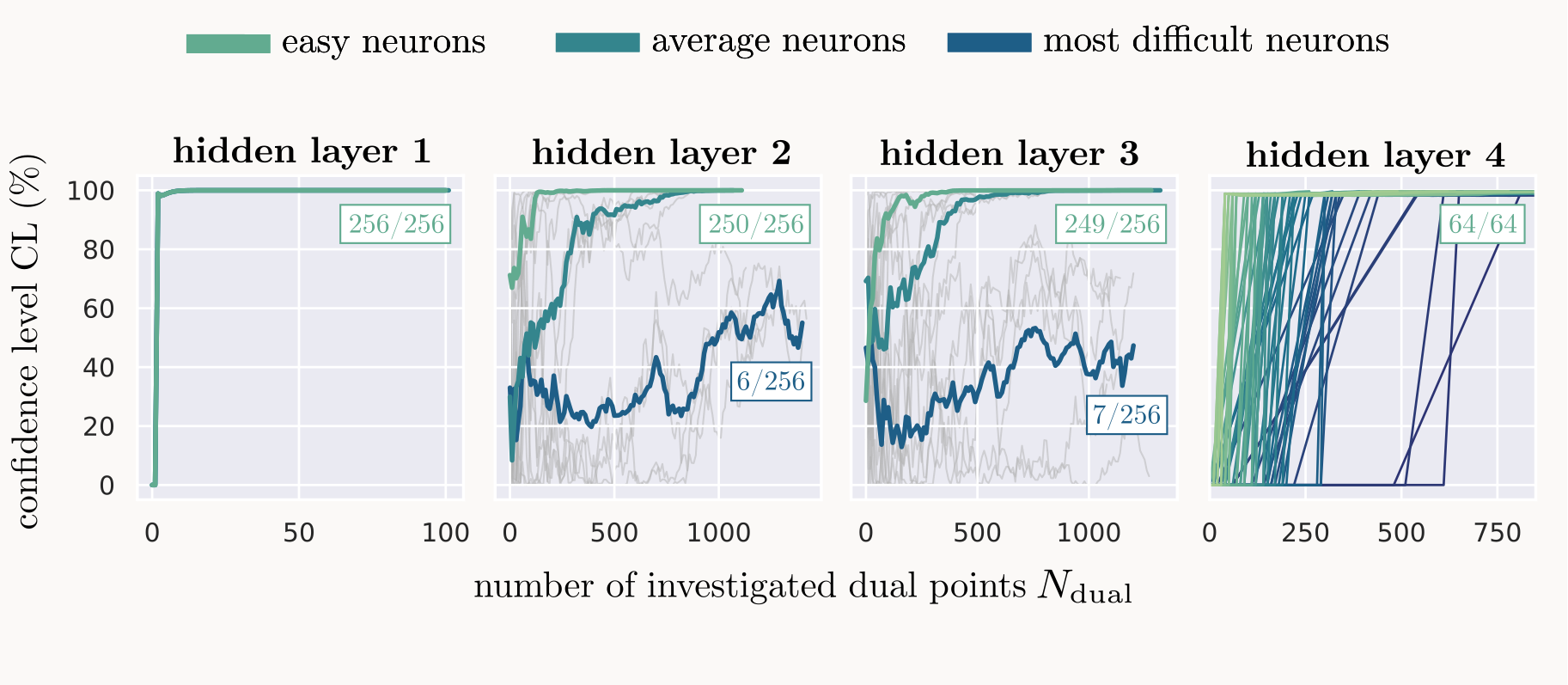}
    \caption{The evolution of the sign recovery confidence level $\CL$ with the investigated number of dual points $\ndual$ for easy neurons, average neurons, and the most difficult neurons in each layer.}
    \label{fig-sign-recovery-results}
\end{figure}

\end{document}